\documentclass[11pt,letterpaper]{article}
\usepackage{jheppub}
\usepackage{amsmath,amssymb,amsfonts,graphicx,tensor}
\usepackage{times,palatino,newpxmath,courier}
\usepackage{subcaption}
\usepackage{cleveref}
\usepackage[dvipsnames]{xcolor}

\newcommand{\K}{\kappa}

\newcommand{\pd}{\rho_\Sigma}
\newcommand{\pp}{p_\Sigma}
\newcommand{\dd}{\mathrm{d}}
\usepackage{enumitem}

\title{Singular hypersurfaces and thin shells in cosmology}
\author{Abhisek Sahu}
\affiliation{Department of Physics and Astronomy, University of British Columbia,\\
6224 Agricultural Road, Vancouver, B.C., V6T 1Z1, Canada}

\emailAdd{abhi@phas.ubc.ca}

\abstract{ 

We analyse spherically symmetric spacetimes obtained by gluing a cosmological region to a Schwarzschild black hole across a singular co-dimension one hypersurface.
Assuming an arbitrary homogeneous and isotropic cosmology, and working in spacetime dimensions greater than three with general cosmological constant, we derive the stress-energy tensor required on the hypersurface directly in terms of the cosmological energy density.
This general framework yields a new exact solution in four dimensions describing a radiation-filled cosmology matched to vacuum through a pressureless dust shell.
A systematic exploration of parameter space reveals twenty-two distinct families of solutions, including bubble-of-cosmology and Swiss-cheese spacetimes with different global and causal structures.
We also discuss possible generalisations of the construction and explain why such thin-shell cosmologies are of interest in the context of holography and quantum cosmology. For negative cosmological constant, a subset of these solutions admits a Euclidean continuation compatible with a holographic interpretation developed in related work.
In addition, we provide a pedagogical introduction to hypersurfaces in general relativity and a practical framework for constructing thin-shell spacetimes.

}

\begin{document}

\maketitle

\section{Introduction}

The proper formulation of junction conditions at hypersurfaces of discontinuities is a fundamental problem in general relativity, with a rich history of work dating back to Sen \cite{Sen_1924_Junctions}, Lanczos
\cite{Lanczos_1924_Junction}, and O'Brien and Synge \cite{oBrien_1952_jump}.
These earlier works relied on the smooth matching of coordinates at the surfaces to distinguish the physical discontinuities, which may arise from a discontinuous distribution of matter; for example at the surface of a star.

Israel in his seminal paper \cite{israel_singular_1966} introduced a coordinate invariant formulation for hypersurfaces with discontinuities. 
In this formulation, the discontinuity of the extrinsic curvature at the junction between two spacetimes is attributed to the presence of matter confined to the hypersurface with some non-zero stress-energy tensor.
Throughout this paper, we shall refer to singular hypersurfaces simply as \textit{thin shells} and the spacetimes formed from joining two different spacetimes at a junction as \textit{thin shell spacetimes}.

When investigating phenomena such as the decay of a false vacuum bubble \cite{coleman1980gravitational, blau_dynamics_1987} or gravitational collapse, one needs to understand the nature of spacetime as it transitions between two distinct spacetimes, with an intermediary region of finite volume.
In such cases, the junction conditions offer a practical means of constructing effective spacetimes by approximating the intermediary regions with a thin shell.
The construction of these thin shell spacetimes is typically more straightforward since the spacetime metric is known on both sides of the shell.
This means that, instead of solving the full Einstein's field equations, one needs to only solve differential equations describing the evolution of the thin shell.

Junction conditions have also become an important tool for constructing low-energy effective gravitational theories in holography. 
Thin shell spacetimes have been used for constructing semi-classical microstates of a black hole \cite{
Balasubramanian_2022_Microscopic,Balasubramanian_2022_microscopicGR, Emparan_2024_Microstates} and de Sitter wormholes \cite{Balasubramanian_2023_deSitter}, understanding negative energy enhancement in holographic CFTs \cite{May_2021_Negative},  
studying bag of gold spacetimes \cite{Fu_2019_Bag}, studying chaos and scrambling \cite{Shenker2013}, and understanding inflation \cite{freivogel_inflation_2006} and cosmology \cite{Sahu2023Bubbles, Antonini_2023_cosmology_random_entanglement, Antonini:2021xar} within AdS/CFT.
Analogous constructions where spacetimes are instead cut off at an `end-of-the-world' hypersurface have been used to describe localisation of gravity on a brane \cite{Karch_2000_Locally, Karch:2000gx}, gravitational duals to boundary conformal field theories \cite{fujita_aspects_2011, Takayanagi2011}, braneworlds \cite{cooper_black_2018, Waddell_2022_Bottom-up} and end-of-the-world black holes \cite{Rozali:2019day, Chen_2020_quantum}.
The above list is of course incomplete and biased by the author's interests.

The Oppenheimer-Snyder solution is a canonical example of a thin shell spacetime and was one of the first models to show how black holes could naturally arise within the framework of general relativity.
In their landmark paper \cite{Oppenheimer1939Continued}, Oppenheimer and Snyder model the interior of a burnt-out massive star as a ball of uniformly distributed pressureless dust with a Friedmann-Lemaître-Robertson-Walker (FLRW) metric. The spacetime outside the star is assumed to be a portion of a Schwarzschild black hole, which joins at the surface of the star without requiring any additional matter. 
Solving Einstein's field equation in this setup, they find an analytical solution where a sufficiently massive star collapses indefinitely, eventually forming an event horizon and becoming causally disconnected from the rest of the universe. 

An important assumption in the Oppenheimer-Snyder solution is neglecting the pressure of the matter within the cosmological region, which is crucial for the cosmological and Schwarzschild regions to be joined without any discontinuity. 
This paper aims to extend the class of solutions exemplified by the Oppenheimer-Snyder solution by considering thin shell spacetimes involving general cosmological spacetimes, in particular those with non-zero pressure \footnote{For related work relaxing different assumptions in spherically symmetric collapse and matching problems, including anisotropic fluids, radiating fluids, or expansion-free motion, see for example
\cite{Herrera:2012uyt,Kumar_radiating_2020} and references therein.}
.

We now present a brief outline of the paper and a summary of the results.

\paragraph{A crash course on hypersurfaces}
In \cref{sec:review}, we discuss some fundamental results regarding hypersurfaces in general relativity. These results include the concepts of intrinsic and extrinsic curvature, the Gauss-Codazzi equations, and Israel's junction conditions. 
None of the results discussed in this section are original. However, we tailor the discussion in a way that lends itself useful for practically constructing thin shell spacetimes.
In particular, we present an algorithm for joining two distinct spacetime regions at a common timelike hypersurface using the junction conditions. We also discuss the intricacies of this procedure and the often overlooked constraints involved. 
The author hopes that this section provides readers with a self-contained and pedagogical introduction to this subject.

\paragraph{Comoving thin shell}
In \cref{sec:Cosmoshell}, we describe the construction of spherically symmetric spacetimes that consist of a portion of a homogeneous and isotropic cosmological spacetime and a portion of a Schwarzschild black hole, connected by a spherical thin shell with a timelike world volume and possibly non-zero surface energy.
Constructing thin shell spacetimes involving cosmologies is challenging because the flux of the stress-energy tensor across an arbitrary hypersurface in the cosmological spacetime is generally not zero. 
This makes it difficult to prevent the matter from leaking out into the vacuum region. Additionally, the shell itself may interact non-trivially with the cosmological spacetime disturbing the homogeneous distribution of matter in the cosmological region.
We address this difficulty by requiring the thin shell to be comoving with the cosmological region. This effectively imposes reflective boundary conditions on the cosmological side of the shell, ensuring that the cosmological and the vacuum regions do not affect one another.
We derive the comoving condition as a natural consequence of the conservation of shell stress energy and the Gauss-Codazzi equations. 

\paragraph{Matter content of the shell}

We model the shell matter as a mixture of ideal fluids that do not interact with each other or the ideal fluids in the cosmological region. The shell remains spherical throughout its evolution with a size varying in time, proportional to the scale factor of the cosmology.
In \cref{sec:shell_matter}, we use Israel's junction conditions to calculate the effective total density and pressure of the shell fluid required for joining the cosmological and vacuum spacetimes, in terms of the scale factor and the energy density of the cosmological matter. 
Apart from the assumption of spherical symmetry, our analysis applies to FLRW with arbitrary fluid density, spatial curvatures, cosmological constants and spacetime dimensions greater than three. 

\paragraph{Radiation-filled cosmologies and shells of dust}

In \cref{sec:dust_radiation} we study the case of cosmologies filled with cold dust and radiation. 
We recover the Oppenheimer-Snyder solution in all spacetime dimensions greater than three in the limiting case with vanishing radiation density. 
In addition, we also find a new exact solution, specific to four spacetime dimensions, with a shell of dust separating the black hole region and the cosmological region filled with dust and radiation.

The free parameters in our set of solutions are the dust and radiation densities and the size of the shell at some initial time slice. 
The length scale of the problem is set by the cosmological constant which is found to be the same on both sides of the shell.
While the ADM mass of the black hole depends on all the free parameters, the density of dust on the shell is entirely determined by the density of radiation. 
Based on the sign of the cosmological constant and whether the cosmological region has a finite size, we find twenty-two distinct families of solutions in \cref{sec:catalogue}, as shown in \cref{fig:TimeSym} and \cref{fig:Mono}.
In \cref{sec:parameter} we explore our three-dimensional parameter space and find that each of these families has a large parameter subspace available to it.

\begin{figure}
    \centering
    \begin{subfigure}[b]{0.33\textwidth}
    \centering
    \includegraphics[width=\textwidth]{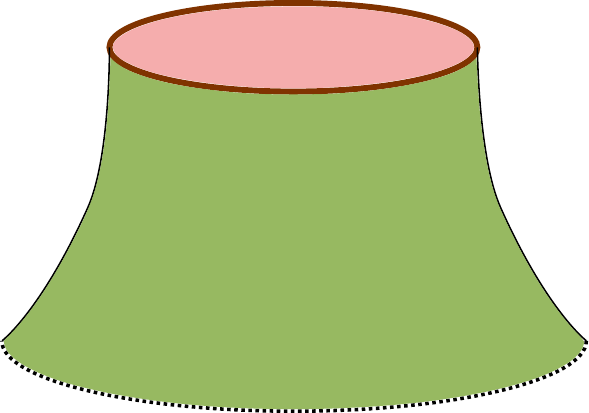}
         \caption{}
         \label{fig:BubbleSlice}
     \end{subfigure}
     \hfill
    \begin{subfigure}[b]{0.48\textwidth}
    \centering
    \includegraphics[width=\textwidth]{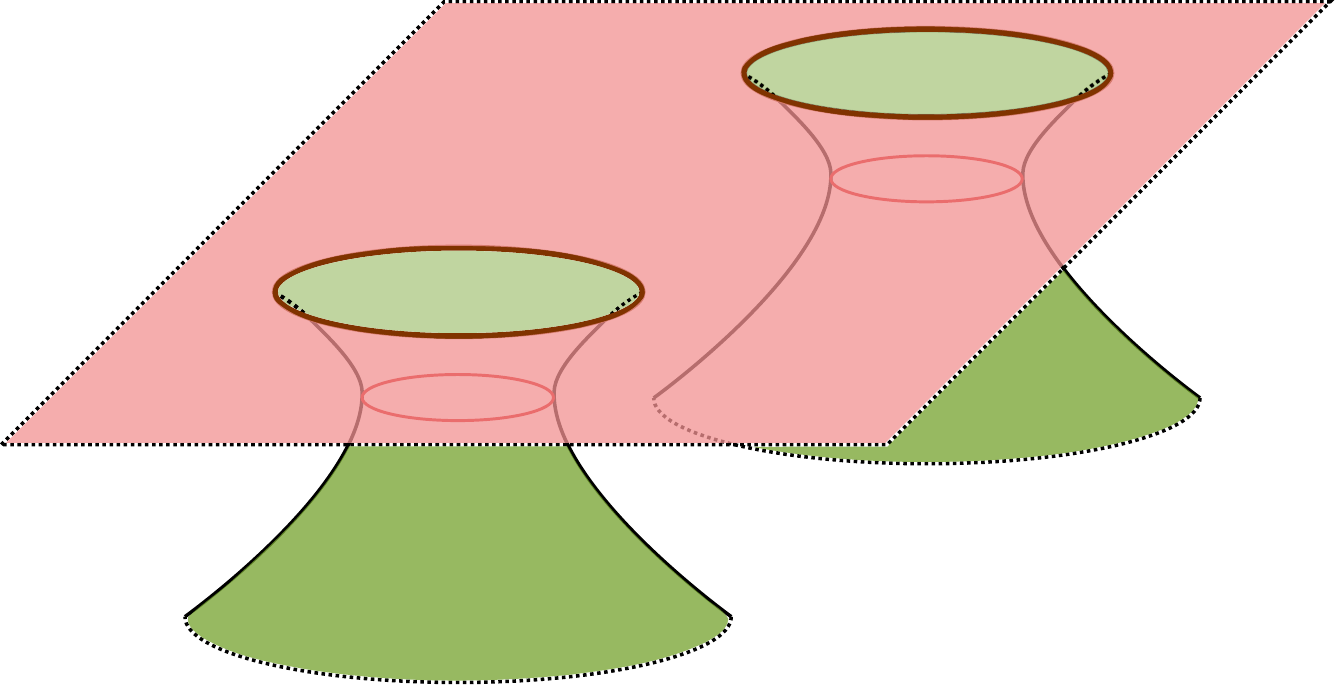}
    \caption{}
     \label{fig:SwissSlice}
     \end{subfigure}
     
     \caption{A sketch of spatial slices of (a) bubble of cosmology spacetime and (b) Swiss cheese spacetime. The cosmological part is shaded in pink, the black hole patch in green, and the thin shell in brown. The thin shell depicted here as a circle is a $(D-1)$-sphere with the other angular dimensions suppressed.}
     \label{fig:SpatialSlice}
\end{figure}

\paragraph{Cosmological bubbles and Swiss cheese}
All the possible solutions we found can be broadly classified into two types (shown in \cref{fig:SpatialSlice}) based on whether the cosmological region is a ball of finite comoving radius, extending infinitely with a finite-sized spherical void. 
The solutions of the former kind, of which the Oppenheimer-Snyder model \cite{Oppenheimer1939Continued} is an example, are referred to as `bubble of cosmology' spacetimes. 

Alternatively, the cosmological part can be infinitely large with arbitrarily many finite-sized vacuoles distributed throughout giving the spacetime an appearance of Swiss cheese, a construction initially discussed by Einstein and Straus \cite{Einstein_1945_influence}.
The vacuoles connect to an event horizon hiding a causally disconnected patch. 
Note that, unlike in the bubble of cosmology spacetime, the Swiss cheese spacetimes can have multiple vacuoles due to the homogeneity of the cosmological spacetime.

\section{Review of thin shell spacetimes}
\label{sec:review}
For the sake of completeness, we briefly review some results from differential geometry concerning hypersurfaces and junction between two spacetimes at a given hypersurface\footnote{For a more complete review the reader is encouraged to refer to \cite{israel_singular_1966, Misner_Gravitation_1973, Poisson2004}}.
A salient feature of the formalism introduced subsequently is that although we use a specific coordinate system to specify the embedding, the final results shall be coordinate-independent. 
We shall use Latin letters to denote tensors on the hypersurface and Greek letters for tensors on the global manifold. Some of the quantities may also carry both Latin and Greek indices.

\subsection{Hypersurfaces in general relativity}

Consider a $D$-dimensional hypersurface $\Sigma$ embedded within a $(D+1)$-dimensional manifold $\mathcal{M}$.
We describe this embedding either by specifying the global coordinates $x^\alpha(y^a)$ as functions of arbitrarily chosen intrinsic coordinates $y^a$ of the hypersurface, or by specifying some constraint $\Phi(x^\alpha) = 0$. 
Infinitesimal displacements \textit{on} the hypersurface are spanned by $D$ mutually orthogonal tangent $(D+1)$-vectors $\mathbf{e}_i$ which are given by
\begin{equation}
\label{eq:tangent_vec_def}
    e^\mu_{i} = \frac{\partial x^\mu(y)} {\partial y^i}.
\end{equation}
The length of a displacement on the hypersurface is determined by the induced metric $h_{ij}$, which is a projection of the bulk metric $g_{\mu\nu}$ onto the hypersurface,
\begin{equation}
    h_{ij} = g_{\mu\nu}e^\mu_{i}e^\nu_{j}.
\end{equation}
Note that $h_{ij}$ behaves like a scalar under general coordinate transformations of the global spacetime.
However, it transforms like a rank 2 tensor under transformations of the intrinsic coordinates. 
Thus the induced metric is a $(D+1)-$scalar, but a $D$-tensor.

The normal vector $\mathbf{n}$ represents displacements \textit{out} of the surface and is orthogonal to all the tangent vectors at any point on the hypersurface.
$\mathbf{n}$ is normalized to $\epsilon(\mathbf{n}) = n_\mu n^\mu = 1$ for time-like surfaces and $-1$ for space-like surfaces\footnote{In this paper we shall only be concerned with timelike hypersurfaces. A formalism for describing null surfaces can be found in \cite{Poisson2004}}.
On the hypersurface, we can decompose the full metric into the tangent and the normal components as
\begin{equation}
g^{\alpha\beta}=h^{ij}e^\alpha_{i}e^\beta_{j} + \epsilon(\mathbf{n})n^\alpha n^\beta.
\end{equation}
The induced metric introduces the notion of an \textit{intrinsic 
connection} defined as
\begin{equation}
\label{eq:intrinsic_connection_def}
    \Gamma_{cab} \equiv \mathbf{e}_c\cdot  \mathbf{e}_{a,b} = g_{\alpha\gamma} e^{\gamma}_{c} e^\beta_b  e^\alpha_{a;\beta},
\end{equation}
where the dot product is defined using the global metric $g_{\mu\nu}$. The subscript `$,$' denotes a partial derivative and `$;$' denotes a global covariant derivative. 

Intrinsic quantities such as the connection \eqref{eq:intrinsic_connection_def}, the intrinsic curvature and the intrinsic covariant derivative can be calculated from the metric $h_{ab}$, without referring to the surrounding manifold $\mathcal{M}$.
The information about the embedding of the hypersurface is contained in the \textit{extrinsic curvature}, which roughly quantifies how the hypersurface folds within the global spacetime. 
It is defined as 
\begin{equation}
\label{eq:extrinsic_curvature_def}
    K_{ab} \equiv \mathbf{e}_a\cdot \mathbf{n}_{,b} = e^\alpha_a e^\beta_b  n_{\alpha;\beta} = -\mathbf{n}\cdot\mathbf{e}_{a,b}, 
\end{equation}
which is nothing but the projection of the covariant derivative of the normal vector along the hypersurface. 
Since ${e}^\mu_{a,b} = {e}^\mu_{b,a}$, as is evident from the definition of the tangent vectors \cref{eq:tangent_vec_def}] $K_{ab}$ is a symmetric tensor.
Furthermore, as ${n}^\mu{n}_\mu = 0$ the partial derivative of the normal vector is purely tangential to the hypersurface, given by
\begin{equation}
\label{eq:n_derivative}
    {n}^\mu_{,b} = \tensor{K}{^a_b} {e}^\mu_a.
\end{equation}
Similarly, we can decompose the partial derivatives of the tangent vector along the normal and tangential components using Eqs. \eqref{eq:intrinsic_connection_def} and \eqref{eq:extrinsic_curvature_def} as
\begin{equation}
\label{eq:gauss_weingarten}
  {e}^\mu_{a,b}  = -\epsilon(\mathbf{n}) {n^\mu}K_{ab} + {e}^\mu_c \Gamma^c_{ab}.
\end{equation}
The above relation is known as the \textit{Gauss-Weingarten equation}.

{
The induced metric $h_{ab}$ and the extrinsic curvature $K_{ab}$ are often referred to
in the literature as the \emph{first} and \emph{second
fundamental forms} of the hypersurface, respectively. The first fundamental form
$h_{ab}$ encodes intrinsic distances measured along $\Sigma$, while the second
fundamental form $K_{ab}$ characterises how $\Sigma$ is embedded in the ambient
spacetime through the normal derivative of the metric. 
}

We can now characterise how tensors `confined' to the hypersurface vary over the surface
\footnote{Such an object may be created, for example, by projecting an arbitrary tensor $\tensor{T}{^{\alpha\dots}_{\beta\dots}}$ defined on $\Sigma$ using the projector $h^{\alpha\beta} = h^{ab}e^{\alpha}_a e^{\beta}_b$. Under projection operation $\tensor{h}{^{\mu}_{\alpha}}\dots \tensor{h}{_{\nu}^{\beta}}\dots\tensor{T}{^{\alpha\dots}_{\beta\dots}}$, only the tangential components survive. 
}.
For example, consider a vector field $A$ that is purely tangential to the hypersurface. Such a $(D+1)$-vector is associated with a $D$-vector whose components are denoted as $A^c$, each of which transforms as a scalar in the global manifold. Then the components of the $(D+1)$-vector $\mathbf{A}$ can be written as $A^\mu = A^c e^\mu_c$ everywhere on the hypersurface. 
We calculate its directed derivative along a tangent to the hypersurface:
\begin{align}
e^\mu_a A^\alpha_{;\mu} &= 
    e^\mu_a  (A^c e^\alpha_c)_{;\mu}\\
    &= A^c e^\mu_a  e^\alpha_{c;\mu} + e^\alpha_c e^\mu_a A^c_{;\mu} \\
    &= -\epsilon(\mathbf{n}) n^\alpha K_{ac}A^c  + e^\alpha_c (A^b \Gamma^c_{ba} + \partial_a A^c) \\
    &=  -\epsilon(\mathbf{n}) n^\alpha K_{ac}A^c  + e^\alpha_c  A^c_{|a}. 
\end{align}
In the third line, we used the Gauss-Weingarten equation \eqref{eq:gauss_weingarten} and also the fact that $A^c$ is a global scalar. 
In the fourth line, the subscript $|a$ denotes an \textit{intrinsic covariant derivative}.
We see that the normal component of $\mathbf{A}_{,a}$ depends on the extrinsic curvature intrinsic, whereas the tangential component, which is simply the intrinsic covariant derivative of $A_a$, depends on the intrinsic curvature.

\paragraph{Gauss-Codazzi equations}
In general relativity, initial value problems are formulated by considering an initial Cauchy slice with $h_{ab}$ and $K_{ab}$ as inputs on this slice. 
These tensors can not be chosen freely however; they satisfy certain constraints called Gauss-Codazzi equations, which we derive subsequently.

Taking a derivative of the Gauss-Weingarten equation \eqref{eq:gauss_weingarten} we obtain,
\begin{align}
\label{eq:e_double_der}
    e^\mu_{a,bc} 
    &= - \epsilon(\mathbf{n})n^\mu(K_{ab,c}+\Gamma^{i}_{ab}K_{ci}) + e^\mu_j (\Gamma^j_{ab,c} + \Gamma^j_{ci}\Gamma^i_{ab} + -\epsilon(\mathbf{n})K_{ab}K^j_{c})
\end{align}
where we further used \cref{eq:n_derivative} and \cref{eq:gauss_weingarten} to eliminate any partial derivatives of the tangent and the normal vector.
The L.H.S of \cref{eq:e_double_der} may be further rewritten as
\begin{equation}
\label{eq:e_double_der_lhs}
    e^\alpha_{a,bc} = (e^\alpha_{a;\beta}e^\beta_b)_{;\gamma}e^\gamma_c = e^\alpha_{a;\beta\gamma}e^\beta_b e^\gamma_c + e^\alpha_{a;\beta}e^\beta_{b,c}.
\end{equation}
Under anti-symmetrisation of the $b,c$ indices in \cref{eq:e_double_der_lhs} the first term on the right hand side gives $-\tensor{R}{^\alpha_\mu_\beta_\gamma}e^\mu_a e^\beta_b e^\gamma_c$, whereas the second term vanishes.
Further anti-symmetrising the $b$ and $c$ indices in \cref{eq:e_double_der}, we obtain
\begin{align}
   \tensor{R}{^\nu_\mu_\beta_\gamma} e^\mu_a e^\beta_b e^\gamma_c = \epsilon(\mathbf{n}) n^\nu (K_{ab|c}-K_{ac|b}) + e^\nu_j \big(\epsilon(\mathbf{n})(K^j_c K_{ab} - K^j_b K_{ac}) + \tensor{R}{^j_a_b_c}\big).
\end{align}
The tangential part of the above equation gives us
\begin{equation}
    \label{eq:gauss_Coda_1}
    R_{\mu\nu\alpha\beta}e^\mu_a e^\nu_b e^\alpha_c e^\beta_d = R_{abcd}+\epsilon(\mathbf{n})(K_{ad}K_{bc}-K_{ac}K_{bd}),
\end{equation}
while the normal part gives us
\begin{equation}
    \label{eq:gauss_Coda_2}
    R_{\mu\nu\alpha\beta}n^\mu e^\nu_a e^\alpha_b e^\beta_c = K_{ab|c}-K_{ac|b}.
\end{equation}
Eq. \eqref{eq:gauss_Coda_1} and \cref{eq:gauss_Coda_2} are collectively called the Gauss-Codazzi equations. 
An interesting feature of these equations is that they relate some components of the global Riemann tensor to the intrinsic Riemann tensor. However, the components $ R_{\nu\mu\beta\gamma}n^\nu e^\mu_a n^\beta e^\gamma_c $ cannot be written in terms of surface quantities. 
We may further formulate these constraints in terms of Einstein's tensor $G_{\alpha\beta}$ by contracting the intrinsic indices appearing in the Gauss-Codazzi equations \eqref{eq:gauss_Coda_1},\eqref{eq:gauss_Coda_2}. These contracted forms are
\begin{align}
    \label{eq:gauss_coda_contract1}
    2\epsilon(\mathbf{n})G_{\alpha\beta}n^\alpha n^\beta &= \epsilon(\mathbf{n})(K^2 - K^{ab}K_{ab}) - R^{(D)} \\
    \label{eq:gauss_coda_contract2}
    G_{\alpha\beta}e^\alpha_a n^\beta &= K^b_{a|b}-K_{,a},
\end{align}
where $R^{(D)}$ denotes the intrinsic Ricci scalar of the $D$-dimensional hypersurface. 

We explicitly see that the extrinsic curvature $K_{ab}$ cannot be chosen independently of the induced metric $h_{ab}$ as a consequence of the Gauss-Codazzi equations.
We shall later use the Gauss-Codazzi equations to formulate constraints in the construction of thin shell spacetimes.

\subsection{Gluing spacetimes together}
\label{sec:gluing}
One can create new spacetimes by simply cutting out regions from two spacetimes $\mathcal{M}_1$,$\mathcal{M}_2$ along a common timelike hypersurface $\Sigma$, and then joining them together with a timelike thin shell which typically has a non-zero stress-energy tensor.
The junction conditions provide a relation between the shell's world-volume, stress-energy tensor, and the surrounding geometry.

\paragraph{Junction conditions} To derive the junction conditions, we consider the metrics $g_{1\mu\nu}$ and $g_{2\mu\nu}$, as well as cosmological constants $\Lambda_1$ and $\Lambda_2$ in the spacetimes $\mathcal{M}_1$ and $\mathcal{M}_2$ respectively. The induced metric and extrinsic curvature of the hypersurface approaching the surface from $\mathcal{M}_1$ and $\mathcal{M}_2$ may differ. We denote them by $h_{1ab}$, $K_{1ab}$ and $h_{2ab}$, $K_{2ab}$ respectively. The gravitational effective action in this case is
\begin{align} 
\nonumber \mathcal{S}= \frac{1}{16 \pi G_N}& \bigg[ \int_{\mathcal{M}_{1}} \dd^{D+1} x \sqrt{|g_1|}\left(R_1-2 \Lambda_1\right)+\int_{\mathcal{M}_{2}} \dd^{D+1} x \sqrt{|g_2|}\left(R_2-2 \Lambda_2\right) \\ 
\label{eq:gravAction}
&+2 \int_{\Sigma} \dd^{D} y \left( \sqrt{|h_1|}K_1-\sqrt{|h_2|}K_2 \right) \bigg] + \mathcal{S}_{m};
\end{align} 
where $|h_1|, |h_2|$ are determinants of the respective induced metrics and $K_1, K_2$ are the traces of the extrinsic curvatures.
The first two terms in \cref{eq:gravAction} represent Einstein-Hilbert terms and the third term includes the Gibbons-Hawking-York terms at the shell $\Sigma$.  We have chosen the normal vector on the shell to point from $\mathcal{M}_1$ to $\mathcal{M}_2$, resulting in a negative sign for the extrinsic curvature term $K_2$ in \cref{eq:gravAction}. 
$\mathcal{S}_m$ represents all the matter contributions, including those localised on the shell itself.

Varying the gravitational action \eqref{eq:gravAction} with respect to the induced metric, we obtain the following equations
\begin{align}
\label{eq:JuncCond1}  [h_{ab}]  &= 0 \\
 \label{eq:JuncCond2} [K_{ab}]  &= {8 \pi G_N}\left(S_{ab} - \frac{1}{D-1}h_{ab}S\right),
\end{align}
where $S_{ab}$ is the stress-energy tensor of the surface matter, defined as,
\begin{equation}
    S^{ab} = - \frac{1}{\sqrt{-h}}\frac{\delta \mathcal{S}_{m}}{\delta h_{ab}}\Bigg|_{\Sigma},
\end{equation}
and $S=h^{ab}S_{ab}$ is the trace of the surface stress energy tensor. 
Here and in the following, the square brackets denote the jump $[T]=T_1 -T_2$ across the thin shell going from region $\mathcal{M}_1$ to $\mathcal{M}_2$. 
The two conditions \cref{eq:JuncCond1} and \cref{eq:JuncCond2}
\footnote{Note that the junction conditions \eqref{eq:JuncCond1} and \eqref{eq:JuncCond2} are tensor identities, and hence coordinate independent. In particular, they do not depend on how coordinates in one spacetime join up with the coordinates in the other spacetime.}
are referred to as \textit{Israel's junction conditions} \cite{israel_singular_1966}.
 The first junction condition tells us that the induced metric {(equivalently, the first fundamental form)} must be continuous, implying that observers on either side of the shell must agree on their length measurements on the shell.
 The second condition implies that the extrinsic curvature {(and consequently the second fundamental form)} may be discontinuous due to the matter degrees of freedom localized on the surface of the shell. 
 This discontinuity corresponds to a layer of infinite energy density, which causes a sudden force on any particle crossing the surface.

\paragraph{An algorithm for gluing}

A thin shell spacetime is commonly constructed by following a step-by-step procedure. 
The first step is to utilise the symmetries of the problem to motivate an ansatz for the embedding of a timelike shell and its associated stress-energy tensor.
We can write a Gauss-Codazzi equation \eqref{eq:gauss_coda_contract1} for the shell on each manifold that relates the extrinsic curvature of a hypersurface with the Einstein tensor of the spacetime it is embedded in. 
This along with the second junction condition gives us
\begin{align}
\label{eq:junctionConstraints}
    [G_{\alpha\beta}]e^\alpha_a n^\beta &= 8\pi G_N \tensor{S}{^b_{a|b}} \\
    [G_{\alpha\beta}]n^\alpha n^\beta &= 8\pi G_N\epsilon(\mathbf{n})\Tilde{K}_{ab} S^{ab},
\end{align}
where $\Tilde{K}_{ab} = \frac{1}{2}(K_{1ab}+K_{2ab})$.
The shell stress energy must also satisfy the above constraints, along with those that arise from symmetry considerations, for the junction between two chosen spacetimes to be consistent.

\begin{figure}
    \centering
\includegraphics[width=0.9\textwidth]{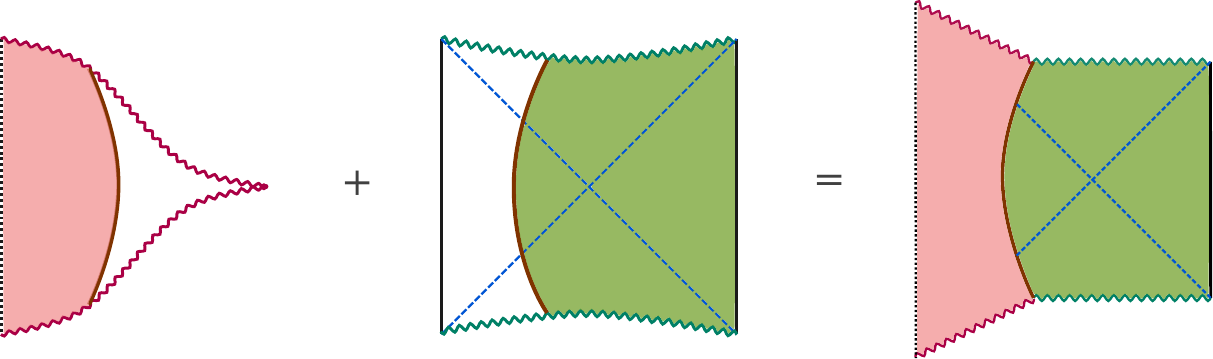}
    \caption{A schematic depiction 
    of the procedure for constructing a thin shell spacetime. 
    A cosmological patch (pink) taken from the left picture is glued to a black hole patch (green) in the middle picture along a thin shell (brown line) to give the resulting thin shell spacetime on the right. 
    {The thin shell moves along a co-moving trajectory in the cosmological spacetime.}
    Note that the Penrose diagram for a thin shell spacetime is only qualitative.  }
    \label{fig:Composite}
\end{figure}

The thin shell divides both $\mathcal{M}_1, \mathcal{M}_2$ into two regions by the shell.
The next step is to determine which of these regions are included in our composite spacetime.
This can be achieved by resolving the sign ambiguity in the normal vector, which is normalized up to an overall factor of $\pm 1$. The different choices correspond to opposite directions. We choose the overall sign of the normal vector such that it points away from the selected region in $\mathcal{M}_1$ and towards the region in $\mathcal{M}_2$.

We assume that the configuration of the shell on an initial time slice is known.
Finally, the junction conditions \eqref{eq:JuncCond1} and \eqref{eq:JuncCond2} offer a set of differential equations that include the embedding functions as well as other functions and parameters related to the shell stress-energy tensor. By solving these differential equations given the initial conditions and parameters of the surrounding spacetime, we can obtain the shell stress-energy tensor and its embedding at any given time.

The algorithm discussed above is schematically depicted in \cref{fig:Composite}.
As an example of this procedure, consider the construction in \cite{freivogel_inflation_2006}, where the authors construct solutions with spherical bubbles of pure de Sitter spacetime embedded within a Schwarzschild-AdS black hole. 
The shell is also assumed to have a constant surface tension and be spherical on every time slice.
The de Sitter region is chosen to have a finite proper size, and a discrete parameter is allowed to choose whether the black hole patch has one or two asymptotic regions.
The junction conditions then determine the evolution of the proper size of the bubble, given the de Sitter cosmological constant,  black hole mass, and the shell tension as inputs.

\section{Thin shells in cosmology}
\label{sec:Cosmoshell}

In this section, we use the algorithm discussed in \cref{sec:gluing} to construct spherically symmetric thin shell geometries with a patch of FLRW spacetime on one side and a vacuum spacetime on the other. Due to Birkhoff's theorem, the vacuum spacetime corresponds to a patch of a Schwarzschild black hole.
We describe both these spacetimes in \cref{sec:cosmoBH} before stating our assumptions for the thin shells and deriving the comoving condition in \cref{sec:thinShell}. Finally, we calculate the density of the shell matter using the junction conditions in \cref{sec:shell_matter}. 

\subsection{The cosmology and the black hole}
\label{sec:cosmoBH}

\begin{figure}
    \centering
    \begin{subfigure}[b]{0.37\textwidth}
    \centering
    \includegraphics[width=\textwidth]{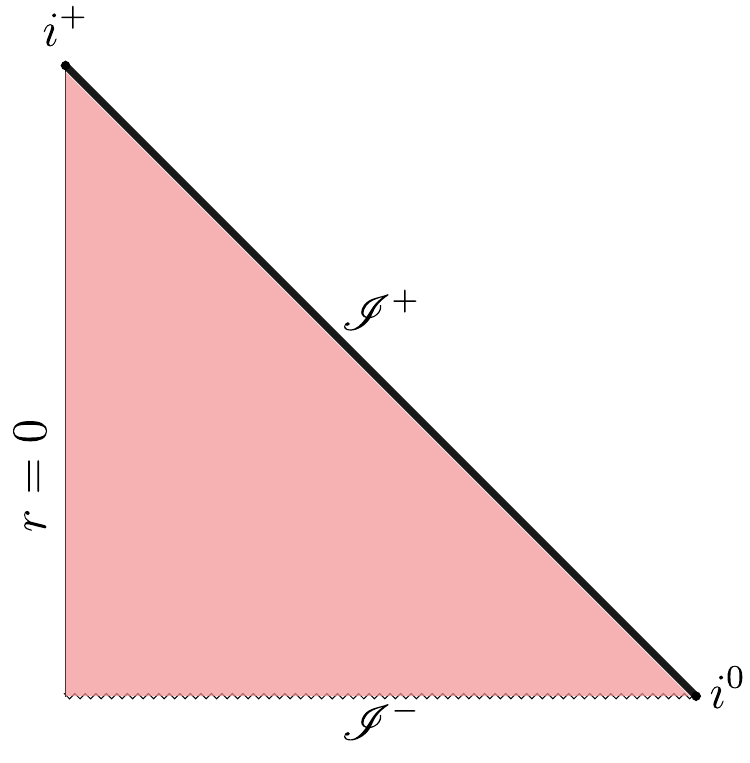}
         \caption{}
         \label{fig:OpenExpanding.pdf}
     \end{subfigure}
     \hfill
    \begin{subfigure}[b]{0.35\textwidth}
    \centering
    \includegraphics[width=\textwidth]{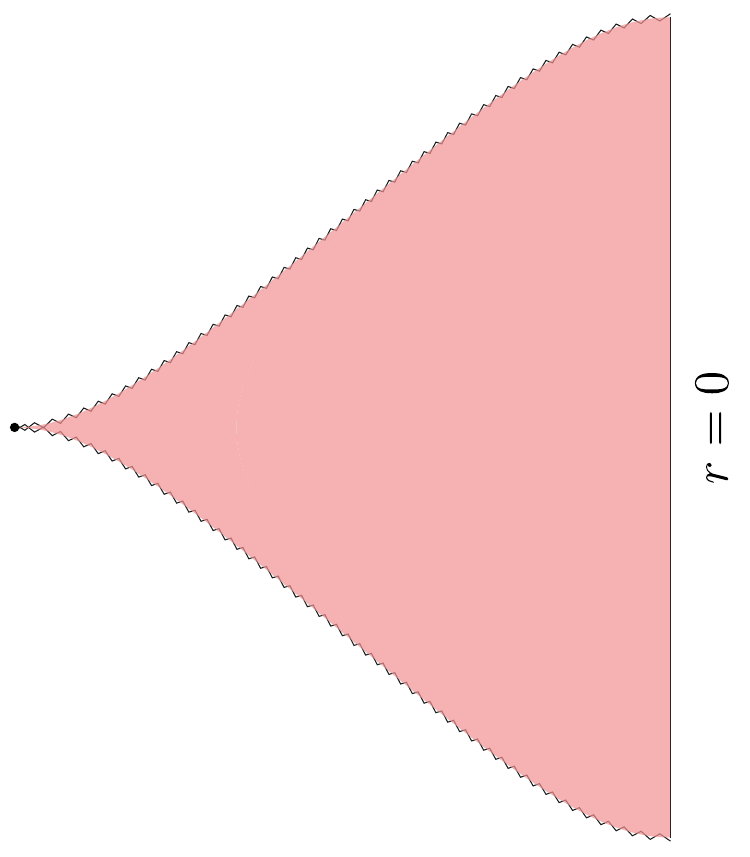}
         \caption{}
         \label{fig:OpenBangCrunch.pdf}
     \end{subfigure}
     \hfill

    \begin{subfigure}[b]{0.4\textwidth}
    \centering
    \includegraphics[width=\textwidth]{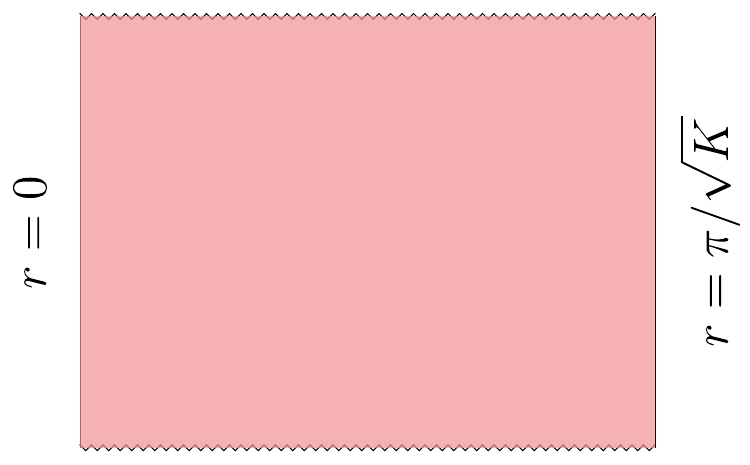}
         \caption{}
         \label{fig:ClosedBangCrunch.pdf}
     \end{subfigure}
     \hfill
    \begin{subfigure}[b]{0.4\textwidth}
    \centering
    \includegraphics[width=\textwidth]{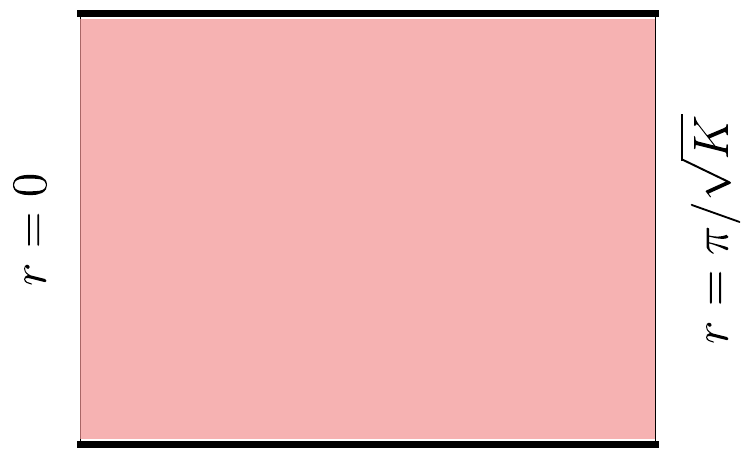}
         \caption{}
         \label{fig:ClosedBounce.pdf}
     \end{subfigure}
    
    \caption{An assortment of cosmological spacetimes: (a) open monotonically expanding cosmology (b)  open time-reversal symmetric cosmology with big bang/big crunch, (c) closed cosmology with big bang/big crunch, and (d) closed bouncing cosmology. The zig-zag lines indicate the big bang/big crunch singularities and the solid black lines represent asymptotic infinities.
        }
    \label{fig:BlackHoles}
\end{figure}

\paragraph{Homogeneous and isotropic cosmology} Region $\mathcal{M}_1$ in our construction is a patch of a FLRW spacetime. In comoving spherical coordinates, the metric of a $D+1$ dimensional FLRW spacetime is,
\begin{align}
\label{eq:FRWMetric}
    ds^2 &= -dt^2 + a^2(t) dr^2 + a^2(t) R_{\K}(r)^2 d\Omega_{D-1}^2, \\
\label{eq:RKr}
    R_{\K}(r) &= \left\{ \begin{array}{ll} \sin(\sqrt{\K} r)/\sqrt{\K} & \K > 0 \cr r & \K=0 \cr \sinh(\sqrt{|\K|} r)/\sqrt{|\K|} & \K < 0 \end{array}, \right. 
\end{align}
where $a(t)$ is a dimensionless scale factor and $\K$ is the Gaussian curvature of the spatial slices. The spacetime is assumed to have a cosmological constant $\lambda$, which has the dimension of inverse length squared. If $\lambda$ is non-zero, the quantity $\ell = \sqrt{1/|\lambda|} $ is defined as the dS (AdS) length when $\lambda$ is positive (negative). Additionally, we assume that the spacetime is filled with a mixture of ideal fluids with a total time-dependent energy density $\rho$ and pressure $p$.
The evolution of the matter density and pressure and the scale factor is determined jointly by the equations
\begin{align}
\label{eq:ContinuityEq}
    \frac{d\rho}{d t} + &D H(t) (p+\rho) = 0, \\
\label{eq:Friedmann1}
    H^2 =&\left( \frac{1}{a}\frac{d{a}}{dt}\right)^2 = \frac{16\pi G_N}{D(D-1)} \left( {\rho} + \frac{\lambda}{8\pi G_N}\right ) -   \frac{\K}{a^2}.
\end{align}
The former equation \eqref{eq:ContinuityEq} is the statement of conservation of stress-energy of the fluid and the second equation \eqref{eq:Friedmann1} is the Friedmann equation.

Often, ideal fluids in cosmology follow a simple equation of state 
\begin{equation}
\label{eq:EqnOfState}
    p = w\rho,
\end{equation}
where $w$ is some real constant. 
In $D+1$ spacetime dimensions,
such an ideal fluid has an energy density $\rho\propto a^{-(D+1)(1+w)}$.
When we have a mixture of several such non-interacting simple fluids the total energy density is a sum of the energy density of individual fluids.

\begin{figure}
    \centering
    \begin{subfigure}[b]{0.3\textwidth}
    \centering
    \includegraphics[width=\textwidth]{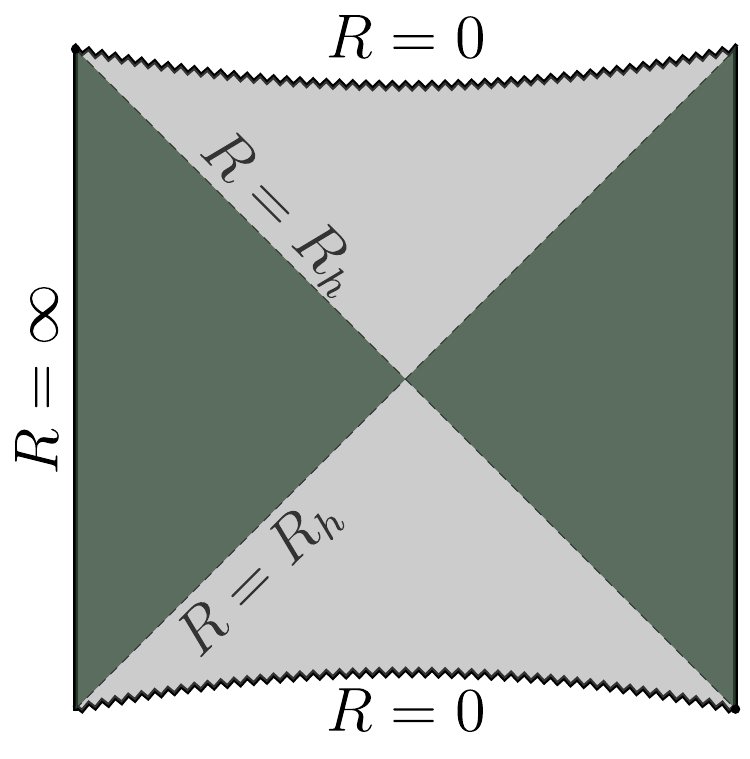}
         \caption{}
         \label{fig:SAdS}
     \end{subfigure}
     \hfill
    \begin{subfigure}[b]{0.5\textwidth}
    \centering
    \includegraphics[width=\textwidth]{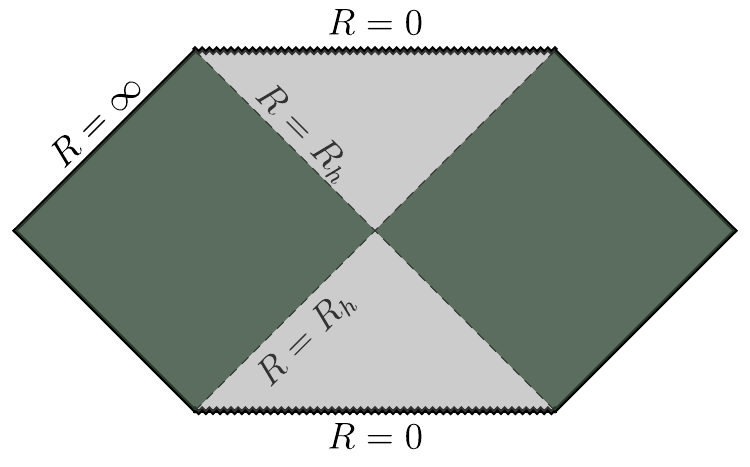}
         \caption{}
         \label{fig:SMink}
     \end{subfigure}
     \hfill

    \begin{subfigure}[b]{0.55\textwidth}
    \centering
    \includegraphics[width=\textwidth]{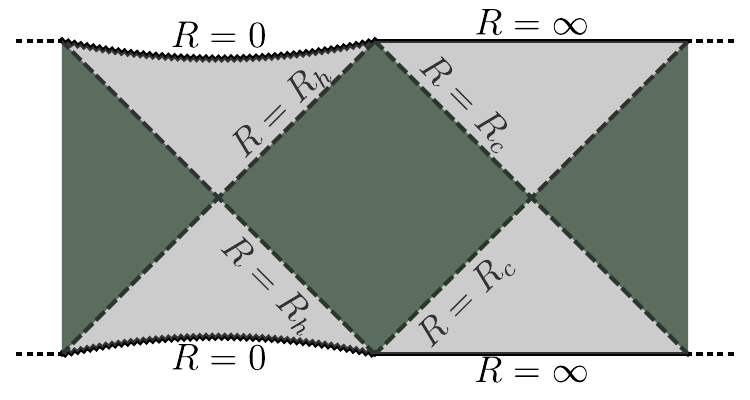}
         \caption{}
         \label{fig:SdS}
     \end{subfigure}
    
    \caption{Penrose diagrams of Schwarzschild black holes in (a) Anti-de Sitter, (b)  Minkowski and (c) de Sitter backgrounds. The zig-zag lines indicate the black hole singularity, the solid black lines represent null infinities, and the dashed lines represent event horizons (as well as cosmological horizons for dS black holes). 
    The maximally extended dS black holes repeat indefinitely in each direction as denoted by the dotted lines.
        }
    \label{fig:BlackHoles}
\end{figure}

\paragraph{Schwarzschild black hole} On $\mathcal{M}_2$ we consider a patch of a Schwarzschild black hole with the metric
\begin{align}
\label{eq:FR}
    dS^2 &= -F(R) dT^2 +\frac{dR^2}{F(R)} + R^2 d\Omega^2_{D-1},
    \\  F(R) &= 1-\frac{2\Lambda}{D(D-1)} R^2- \frac{\mu}{R^{D-2}}.
\end{align} 
$\Lambda$ is the cosmological constant and $\mu$ is a parameter proportional to the black hole mass.
While $\mu$ can take any positive value for AdS and Minkowski black holes,  there is a maximum limit to $\mu$ in the case of dS black holes,
\begin{equation}
    \mu \leq \frac{2}{D-2} \left( \frac{D(D-1)}{2\Lambda} \right)^{\frac{D-2}{2}},
\end{equation}
for a black hole horizon to exist. 
In the above limit, $F(R)$ has two real zeroes. The higher of the two zeroes $R=R_c$ corresponds to a cosmological horizon and the lower zeroes $R=R_h$ corresponds to a black hole horizon.
The causal structures of various Schwarzschild spacetimes are shown in \cref{fig:BlackHoles}.

Note that the Schwarzschild coordinates do not cover the entire maximally extended black hole. 
Two copies of the coordinate patch $R\in(0,R_h), T\in (-\infty,\infty)$ cover the past/future wedge of the black hole interior,  
and two copies of $R\in(R_h,\infty),\, T\in (-\infty,\infty)$ cover the exterior of the black hole.

\subsection{The thin shell}
\label{sec:thinShell}
Let the stress-energy tensor of the thin shell $\Sigma$ be $S_{ab}$. 
We are interested in the simplest scenario where the thin shell does not couple with the cosmological spacetime. Thus we assume that the surface stress energy must be conserved, i.e., 
\begin{equation} 
\tensor{S}{^a_b_{|a}}= 0.
\end{equation}
We shall now prove that the conservation of the surface energy constrains the shell's trajectory to be comoving with respect to the cosmology. 

Recall that \cref{eq:junctionConstraints} obtained from contraction of the Gauss-Codazzi equation gives us 
\begin{equation}
    [G_{\alpha\beta}]e^\alpha_a n^\beta = 8\pi G_N \tensor{S}{^a_b_{|a}} .
\end{equation}
Since the spacetimes on each side satisfy Einstein's equation, we can replace the Einstein tensor with the stress-energy tensor and the cosmological constant term, which implies
\begin{align}
\label{eq:GaussCodazzi2}
    8\pi G_N[T_{\alpha\beta}]e^\alpha_a n^\beta &- [\Lambda g_{\alpha\beta}]e^\alpha_a n^\beta=  8\pi G_N\tensor{S}{^a_b_{|a}} .
\end{align}
The matter in the FLRW region has a stress-energy tensor
\begin{equation}
    T_{\alpha\beta} = (\rho+p)u_\alpha u_\beta + pg_{\alpha\beta},
\end{equation}
where $u^\alpha$ is the streamline velocity of the cosmological fluid.  
On the other hand, $T_{\alpha\beta}=0$ in the black hole region. 
Imposing the conservation of surface stress-energy tensor $\tensor{S}{^a_b_{|a}}$ we obtain, 
\begin{equation}
\label{eq:trajectoryConstraint}
    (\rho+p)(u_\alpha e_{a}^\alpha)(u_\beta n^\beta) = 0,
\end{equation}
where we used the fact that $e$ and $n$ are orthogonal to eliminate the terms containing their dot product. 

Eq. \eqref{eq:trajectoryConstraint} implies $u^\alpha n_\alpha=0$ \footnote{Since $p,\rho$ represent energy sources other than the cosmological constant, $\rho + p \neq 0$. Moreover, $e^\alpha_a u_\alpha \neq 0$, since $u^\alpha$ itself is tangential to the shell and it can not be orthogonal to all the tangent vectors $e^\alpha_a$.}.
In comoving coordinates, $u^\alpha = (1,0,0,0)$, and thus $u^\alpha n_\alpha=n_t=0$.
If the hypersurface is characterised by some function of the coordinates $\Phi(t,r,\Vec{\theta})=0$, $n_t \propto \partial_t\Phi = 0$ implies $\Phi$ is a function of the spatial coordinates only. 
Thus, the shell is comoving with respect to an FLRW observer, i.e., it is a collection of points with fixed comoving coordinates satisfying the condition $\Phi(r,\Vec{\theta})=0$. 
Physically, the comoving shell simply manifests reflecting boundary conditions; any energy flux incident on the shell from the FLRW side is reflected and the energy distribution within the bubble continues to be homogeneous and isotropic.

Secondly, we assume the world volume of the shell is a $(D-1)$-sphere with a time-varying proper size, separating spherically symmetric patches of the FLRW and the black hole spacetime. 
To be consistent with spherical symmetry in the black hole region and homogeneity and isotropy in the cosmological region, the shell matter must also be distributed uniformly and isotropically.  The most general form of the shell matter consistent with this choice is that of a perfect fluid with density and pressure $\rho_\Sigma, p_\Sigma$ respectively, with the stress-energy tensor
\begin{equation}
    S_{ab} = (\pd+\pp)u_a u_b + \pp h_{ab},
\end{equation}

\subsection{Shell matter from cosmological matter}
\label{sec:shell_matter}

Since the shell is spherically symmetric and comoving, let it be a fixed radial coordinate $r=r_0$ from the perspective of an FLRW observer. 
Using Eq. \eqref{eq:FRWMetric}, the induced metric on the shell is 
\begin{equation}
\label{eq:shellMetric}
    ds^2_\Sigma = -dt^2 + a(t)^2R^2_{\K}(r_0) d\Omega_2^2.
\end{equation}
Thus the proper size of the shell is proportional to the scale factor. 
On the black hole side, the embedding of the shell can be described by the coordinates $(T(t), R(t), \Vec{\theta}).$
Once we fix the energy content in the FLRW spacetime, the scale factor and the shell trajectory are fixed up to some constant.
The junction conditions determine the shell stress-energy tensor instead.

The contents of the first junction condition \eqref{eq:JuncCond1} are exhausted by comparing the induced metric for this embedding on the black hole side with Eq. \eqref{eq:shellMetric}. By doing so we obtain,
\begin{equation}
\label{eq:JC1Schwarz}
    R(t) = R_{\K}(r_0)a(t), \quad F(R(t))\dot{T}^2-\frac{\dot{R}^2}{F(R(t))} =1.
\end{equation}
We see that $R(t)$ is simply proportional to the scale factor and $T(t)$ can be found by solving a first order differential equation. 

To apply the second junction condition, we calculate the relevant quantities on both sides of the shell.

\paragraph{Normal vectors and sign conventions}
The unit normal vectors in the FLRW region and the black hole region are given by
\begin{equation}
    n_{\alpha} = q(0,a(t),\Vec{0}), \quad  N_{\alpha} = Q(-\dot R, \dot T, \Vec{0}), \quad q,Q = \pm 1,
\end{equation}
where the over dot represents a $t$-derivative, and the discrete variables choose the cosmological and black hole regions included in the composite spacetime.
Recall that the normal vector points away from the FLRW spacetime region into the Schwarzschild region. To retain the $r<r_0$ region of the FLRW spacetime, the normal vector on the shell must point towards a region with $r>r_0$, which corresponds to the choice $q=1$. On the other hand, $q=-1$ corresponds to the retention of the region with $r>r_0$. 
In the Schwarzschild spacetime, the shell is completely behind a horizon \footnote{When the shell is in the exterior of the black hole horizon ($F(R)>0$), we have the option to choose whether or not the retained region includes the horizon at a constant $T$ slice. If $\dot T > 0$, the normal points towards spheres of higher $R$ if $N^R > 0$, hence, the included region does not contain the horizon. In this case, $Q=1$. Conversely, when $Q=-1$, a constant $T$ slice of the included region intersects the horizon.}.  when $Q=-1$ and outside the horizon for some part of its trajectory for $Q=1$. We shall allow $q, Q$ to be free now and return to fixing them while cataloguing different types of solutions in \cref{sec:catalogue}.

\paragraph{Extrinsic curvatures}
In the FLRW geometry the extrinsic curvatures\footnote{Note that $K^{\theta_j}{}_{\theta_j}$ is the same for all of the $D-1$ angular coordinates, due to spherical symmetry. For conciseness, we shall use $\theta$ to mean any $\theta_j$.  } are given by
\begin{align}
     \tensor[]{K}{_1^{\theta} _{\theta}} &= - h^{\theta \theta} \Gamma^r_{\theta \theta} n_r = {\frac{1}{2}} h^{\theta \theta} g^{rr} \partial_r g_{\theta \theta} n_r =  \frac{q R_{\K}'(r_0)}{R_{\K}(r_0) a(t)}, \\
     \tensor[]{K}{_1^{t} _{t}} &=  - h^{t t} \Gamma^r_{t t} n_r = {\frac{1}{2}} h^{t t} g^{rr} \partial_r g_{t t} n_r = 0.
\end{align}
In the Schwarzschild spacetime the $^{\theta}{} _{\theta}$ component of the extrinsic curvature is calculated as follows:
\begin{align}
    \tensor[]{K}{_2^{\theta} _{\theta}} &= - h^{\theta \theta} \Gamma^R_{\theta \theta} N_R = {\frac{1}{2}} h^{\theta \theta} g^{RR} \partial_R g_{\theta \theta} N_R = \frac{\beta}{R_{\K}(r_0)a(t)} \\
    \label{eq:betaFormula}
    &\beta = Q\sqrt{F(R_{\K}(r_0)a(t))+R_{\K}(r_0)^2 \dot{a}^2}.
\end{align}
To calculate the $^t{}_t$ component, we observe that 
\begin{equation}
    K_{2tt} = \nabla_{\mu}(N_{\nu}u^{\nu})u^{\mu}-N_{\nu}u^{\mu}\nabla_{\mu}u^{\nu} = -N_{\nu}A^{\nu} = -N_T A^T - N_R A^R,
\end{equation}
where $u^\nu = e^\nu_t$ is the proper velocity of an observer sitting at some fixed angular coordinates on the shell and $A^{\nu}$ is the corresponding acceleration.
As $K_{2tt}$ measures the normal component of the acceleration of a particle moving along with the shell, the jump in the extrinsic curvature reflects a change in acceleration of a particle due to a sudden force felt by a particle while crossing the shell.
In the Schwarzschild coordinates the components of $\mathbf{u}$ are $(\dot T, \dot R, 0)$.
The acceleration is given by  $A^{\nu} = \dot u^\nu + \Gamma^{\nu}_{\alpha\beta}u^\alpha u^\beta$ in terms of the Christoffel symbols. 
But we know, $u_{\nu} A^\nu = 0$ and $u^\nu N_\nu = 0$, giving us $A^T = - u_{R} A^R/ u_{T}.$
This gives us
\begin{align}
    \tensor[]{K}{_2_t_t} &= \frac{A^R}{u_{T}}(N_T u_{R} - N_R u_{T}) = \frac{Q A^R}{u_{T}}.
\end{align}
$A^R$ can be calculated using the Christoffel symbols in the Schwarzschild spacetime. It evaluates to 
\begin{align}
\label{eq:ARformula}
    A^R &= \ddot{R} + \frac{1}{2} F(R(t)) \partial_R F(R(t)) \dot{T}^2 - \frac{\partial_R F(R(t))}{2F(R(t))} \dot{R}^2 \\
       &= \frac{1}{2} ( \ddot{R} + \frac{\partial_R F(R(t))}{2} ) = \frac{\beta \dot{\beta}}{\dot{R}},
\end{align}
where in the second line we used \cref{eq:JC1Schwarz} and the definition of $\beta$ from \cref{eq:betaFormula}.
Using \cref{eq:ARformula} and the fact that $u_T = -F(R)\dot{T} = -\beta$ we obtain, 
\begin{equation}
    \tensor[]{K}{_2^t_t} = \frac{\dot \beta}{R_{\K}(r_0)\dot{a}(t)} \;.
\end{equation}

\paragraph{Imposing second junction condition}
{We have now obtained all the non-zero components of the extrinsic curvatures on the cosmological side as well as the black hole side, which we can plug into the second junction condition.
Recall that the second junction condition is 
\[
K_{1ab} - K_{2ab}  = {8 \pi G_N}\left(S_{ab} - \frac{1}{D-1}h_{ab}S\right),
\]
and the stress energy of the shell is given by 
\(S_{ab} = (\pd+\pp)u_a u_b + \pp h_{ab} \).
The trace of the shell stress energy tension is $S = (\rho_\Sigma + p_\Sigma)(-1) + p_\Sigma(D) = p_\Sigma(D-1) -\rho_\Sigma$.
Plugging these expressions into the second junction condition,} \cref{eq:JuncCond2},  the 
 $^{\theta} {} _{\theta}$ component  gives us \footnote{From here on, we will absorb the factor of $8\pi G_N$ in the $\pd$ and $\pp$ for the sake of brevity.}
\begin{align}
\label{eq:JC2applied}
    &\frac{qR'_\K(r_0)-\beta}{R_{\K}(r_0)a(t)} = \frac{1}{D-1}\pd
\end{align}
{
and the $^t{}_t$ component gives us
}
\begin{equation}
\label{eq:JC2applied2}
    \frac{\dot\beta}{R_{\K}(r_0)\dot{a}(t)} = \frac{D-2}{D-1}\pd + \pp.
\end{equation}
{

By solving for $\beta$ from the first equation and differentiating, one can check that $\pd$ and $\pp$ satisfy the continuity equation \eqref{eq:ContinuityEq}.

Using the expression of $\beta$ from Eq. \eqref{eq:betaFormula}, and substituting in \cref{eq:JC2applied} we obtain
\[
\pd = (D-1)\left( \frac{qR'_\K(r_0)}{R(t)} - Q\sqrt{\frac{F(R(t))}{R(t)^2} + \frac{R_\kappa(r_0)^2\dot{a}(t)^2}{R_\kappa(r_0)^2 a(t)^2}} \right).
\]
Note that under square root, we have a term $(\dot a/a)^2$, which is nothing but $H^2$ appearing in the Friedmann equation \eqref{eq:Friedmann1}.
Thus, substituting the Friedmann equation  and the expression of $F(R)$ from \cref{eq:FR} we obtain}
\begin{equation}
\label{eq:shellfromBulk}
    \pd = (D-1)\left( \frac{qR'_\K(r_0)}{R(t)} - Q\sqrt{\lambda-\Lambda + \Big(\frac{R'_\kappa(r_0)}{R(t)}\Big)^2 + \rho - \frac{\mu}{R(t)^{D}}} \right).
\end{equation}
Note that we also use \cref{eq:RKr}, and note that $R'_\kappa(r_0)^2 = 1-\kappa R_\kappa(r_0)^2$.

In the expression above, the quantities on the right hand side include the energy density of matter, curvature and the scale factor in the FLRW region, as well as the black hole mass and the cosmological constant on either side. 
Thus Eq. \eqref{eq:shellfromBulk} calculates the energy density of the matter on the shell as a function of parameters of the spacetime on either side of it. 
The shell pressure $\pp$ can be calculated using $\pd$ and the conservation equation \cref{eq:ContinuityEq} for a $D$ dimensional spacetime.
In principle, the time dependence of $\pd$ and $\pp$ can be removed to obtain an equation of state $\pd(\pp)$.

\section{Dust and radiation filled cosmology}
\label{sec:dust_radiation}

In this section, we consider cosmological spacetime filled with pressureless dust and radiation and provide some explicit examples of thin shell spacetimes where the shell fluid density is proportional to the pressure. 
In addition, we study the parameter space of our solutions and categorise them into distinct families.

The energy density in a dust and radiation filled cosmology takes the form
\begin{equation}
\label{eq:dustAndRad}
    \rho(t) = \frac{\rho_M}{a(t)^{D-1}} + \frac{\rho_R}{a(t)^{D}},
\end{equation}
where $\rho_M$ and $\rho_R$ are some proportionality constants setting the density of dust and radiation at some initial slice where $a=1$.
We can understand the causal structure of these cosmologies by understanding the resulting family of scale factors.
Re-writing the Friedmann equation \cref{eq:Friedmann1} as
\begin{equation}
    \dot{a}^2 + U(a) = 0, \quad U(a) = \kappa-\frac{\rho_M}{a}-\frac{\rho_R}{a^2}-\lambda a^2,
\end{equation}
the scale factor can be thought of as the trajectory of a particle moving in a potential well $U(a)$ with zero total energy. The possible shapes of $U(a)$ for different parameters allow monotonic scale factors as well as scale factors with just one turning point. 

In some cases (discussed below) $U(a)$ has one or more zeroes resulting in a moment of $Z_2$ symmetry with $\dot{a}$ vanishing here. 
Then the scale factor may be zero to start with (big bang), reach a maximum value and decrease to zero in a time-reversal symmetric way to give a big crunch singularity.  
Alternatively, we can have a big bounce cosmology where the scale factor attains a minimum value at the time symmetric slice, and increases to infinity in some finite proper time.
For $\lambda >0$ cosmologies  $U(a)$ has two zeroes when the curvature $\K$ is greater than some positive critical curvature $\K_c$. In 3+1 dimensions, $\K_c$ is implicitly given in terms of $\rho_M, \rho_R, \lambda$ by
\begin{equation}        
\label{eq:CriticalK}
\frac{8\K_c^3\left(\sqrt{9\rho_M^2 + 32\K_c\rho_R}+\rho_M\right)}{\left(\sqrt{9\rho_M^2 + 32\K_c\rho_R}+3\rho_M\right)^3} = \lambda.
\end{equation}
The zero with $U'<0$ gives the minimum scale factor for a big bounce cosmology, whereas the one with $U'>0$ corresponds to the maximum scale factor for a big bang/big crunch cosmology. 
In addition to this case, big bang/big crunch cosmologies also occur when $\lambda=0 $and $\K>0$ or 
when $\lambda<0$. 

For values of $\lambda$ and $\K$  other than those mentioned above, the corresponding scale factors are monotonic.
Here we shall only consider monotonically expanding scale factors corresponding to universes beginning at a big bang with zero scale factor. 

\subsection{Shells with simple equations of state}
\label{subsec:SimpleState}

The matter obtained from Eq. \eqref{eq:shellfromBulk} for an arbitrary FLRW spacetime may in general have complicated physics with a non-trivial equation of state. 
In this subsection we investigate if the shell matter can have a simple physical description as  a mixture of non-interacting ideal fluids in $D$ spacetime dimensions, each of which has a simple equation of state \eqref{eq:EqnOfState}. 

Due to the conservation of shell stress-energy tensor, the energy density of each of these fluids is of the form $\rho_m a^{-m}$ where $m$ is some real number and $\rho_m$ is a proportionality constant. 
For a dust and radiation filled cosmology, the mixture of the fluids must satisfy the condition obtained from \cref{eq:shellfromBulk},  
\begin{equation}
    \left(  \sum_m \frac{\rho_m}{a^m} - \frac{qR_1}{R_0 a}\right) ^2 = \lambda-\Lambda + \left( \frac{R_1}{R_0 a} \right)^2 + \frac{\rho_M R_0^{D} - \mu}{(R_0 a)^{D}} + \frac{\rho_R}{a^{D+1}}, 
\end{equation}
where $R_0=R_{\K}(r_0) $ and $R_1=R_{\K}'(r_0)$.

{Both sides of this equation are polynomials of $1/a(t)$, and hence have to match term by term to be valid at all $t$.
If the LHS is a square of $n$ distinct powers of $1/a$, there will be $n(n+1)/2$ distinct terms. But on the RHS we have $4$ different terms. The two sides can match either when $n=2$ and one of the terms on the RHS is 0, or when $n=1$ and three terms on the RHS are 0. Thus the above equation can be satisfied in only two ways without requiring negative energy density and pressure, mentioned subsequently.
Other configurations are also possible; for example in $D=4$, setting  $\lambda>\Lambda$ and adjusting $\mu=\rho_M R_0^3$. In this case, the shell has a superposition of dust, constant tension and a fluid with $w=-1/2$.
}

    \paragraph{Empty shell:} For any dimension $D\geq 3$ when $\rho_R=0$, $\lambda=\Lambda$ and the mass of the black hole is
     \begin{equation}
         \mu = \rho_M R_0^{D},
     \end{equation}
     we have a shell with no energy or pressure on it ($\rho_m =0$). 
    Thus the cosmological spacetime smoothly joins up with the black hole spacetime along the empty shell. Moreover, when the FLRW has flat spatial slices ($\K=0$), we see that the black hole mass $M$ is precisely equal to the energy contained in the FLRW in terms of dust, using $\mu=2GM$. 
     
This solution is valid for any $\lambda$, any kind of scale factor and any choice of $q$ and $Q$ consistent with $\pd$ being 0. 
     The Oppenheimer-Snyder solution is a particular member of this family of geometries with $\lambda=0$ and $q=Q=1$.
    The geometry with $q=-1$ is a Swiss cheese spacetime where the black hole spacetime exists within a spherical region hollowed out of the FLRW spacetime.
     
   \paragraph{Shell of pressureless dust:}In $3+1$ spacetime dimensions ($D=3$), the cosmology can have some radiation along with pressureless dust. 
    However, we must add pressureless dust to the thin shell to compensate for the radiation pressure. 
    Like the previous case, we need $\lambda=\Lambda$ and the black hole mass to be fine tuned to
    \begin{equation}
    \label{eq:muDustAndRad}
        \mu = \rho_M R_0^3 + 2q\sqrt{\rho_R}R'_K(r_0)R_0^2,
    \end{equation}
 giving us the energy density on the shell
    $\pd = \frac{2\sqrt{\rho_R}}{a^2}.$
    This
    corresponds to the energy density of a fluid in $2+1$ spacetime dimensions with $w=0$, which is pressureless dust.
    Because of the dust on the shell, there is a discontinuity in the extrinsic curvature on both sides which results in a discontinuity in the acceleration of a particle crossing the shell. 
    Both bubbles of cosmology and Swiss cheese spacetimes are allowed in this case. 
    
     Interestingly $\pd$ depends only on the radiation energy density $\rho_R$ and we recover the Oppenheimer-Snyder case when $\rho_R=0$. 
     More interestingly, such a solution is only possible in $3+1$ spacetime dimensions, unlike the generalised Oppenheimer-Snyder case which is possible in any dimensions.   
     
\subsection{A catalogue of thin shell geometries}
\label{sec:catalogue}

\begin{figure}
     \centering
     \begin{subfigure}[b]{0.29\textwidth}
        \centering
        \includegraphics[width=\textwidth]{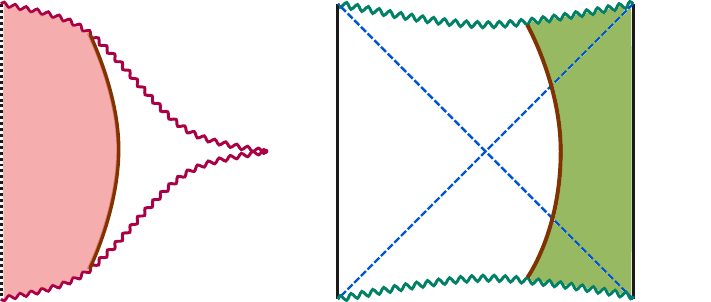}
        \caption*{(A1a)}
        \label{fig:A1a}
     \end{subfigure}
     \hfill
     \centering
    \begin{subfigure}[b]{0.29\textwidth}
         \centering
         \includegraphics[width=\textwidth]{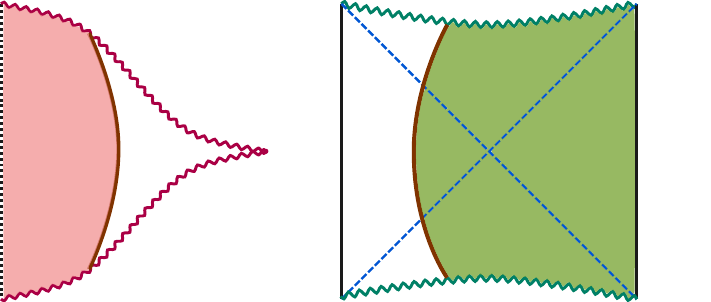}
         \caption*{(A1b)}
         \label{fig:A1b}
     \end{subfigure}
     \hfill
     \centering
    \begin{subfigure}[b]{0.29\textwidth}
        \centering
        \includegraphics[width=\textwidth]{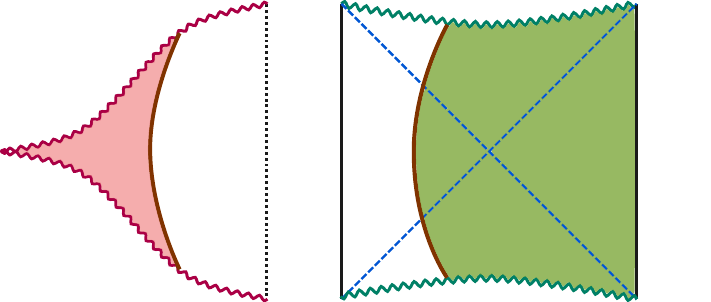}
        \caption*{(A2)}
        \label{fig:A2}
     \end{subfigure}
      
      \vspace{5mm}
      \centering
      \begin{subfigure}[b]{0.29\textwidth}
         \centering
         \includegraphics[width=\textwidth]{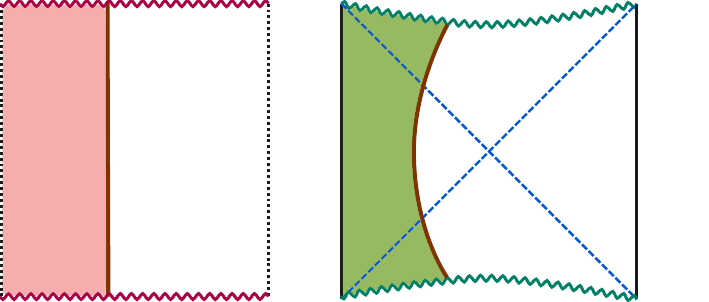}
          \caption*{(A3a)}
         \label{fig:A3a}
     \end{subfigure}
     \hfill
     \centering
     \begin{subfigure}[b]{0.29\textwidth}
         \centering
        \includegraphics[width=\textwidth]{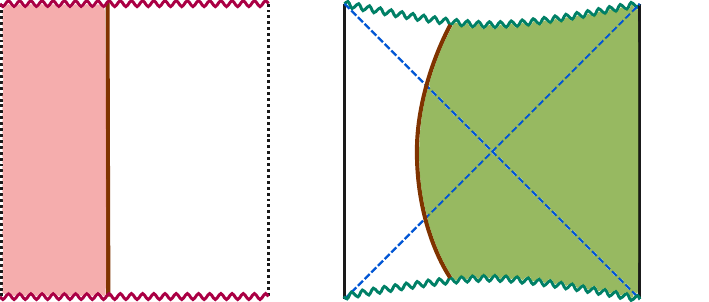}
         \caption*{(A3b)}
         \label{fig:A3b}
     \end{subfigure}
     \hfill
     \centering
     \begin{subfigure}[b]{0.29\textwidth}
         \centering
        \includegraphics[width=\textwidth]{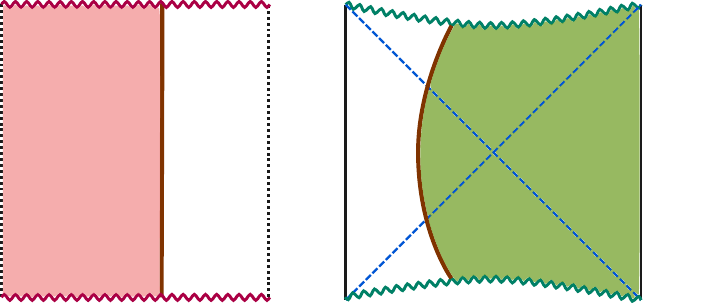}
         \caption*{(A4)}
         \label{fig:A4}
     \end{subfigure}
     
     \vspace{5mm}
     \centering
     \begin{subfigure}[b]{0.29\textwidth}
        \centering
        \includegraphics[width=\textwidth]{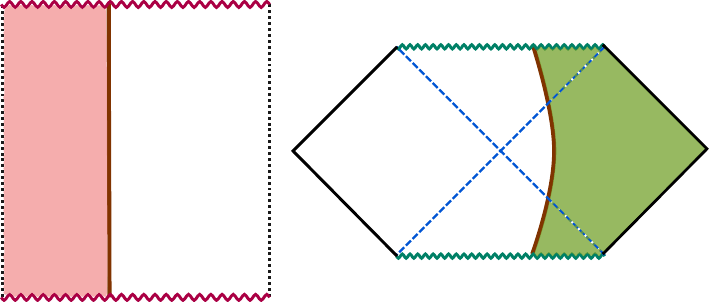}
        \caption*{(M1a)}
        \label{fig:M1a}
     \end{subfigure}
     \hfill
     \centering
     \begin{subfigure}[b]{0.29\textwidth}
        \centering
        \includegraphics[width=\textwidth]{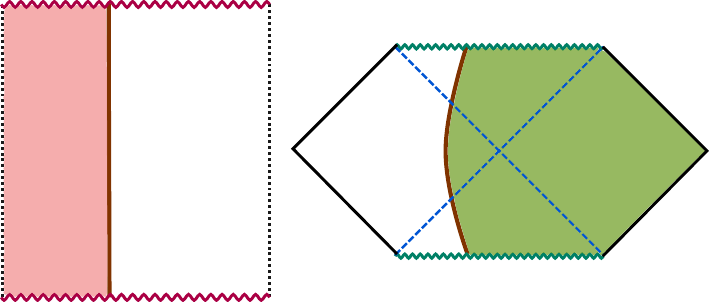}
        \caption*{(M1b)}
        \label{fig:M1b}
     \end{subfigure}
     \hfill
     \centering
     \begin{subfigure}[b]{0.29\textwidth}
        \centering
        \includegraphics[width=\textwidth]{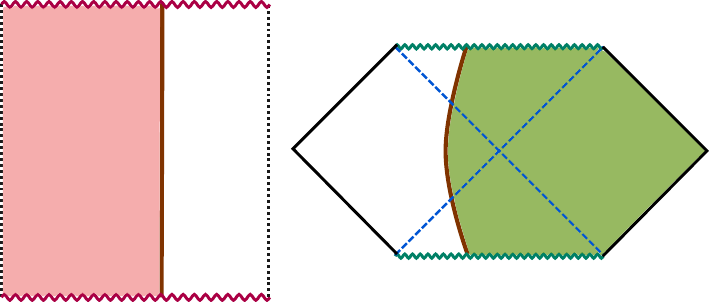}
        \caption*{(M2)}
        \label{fig:M2}
     \end{subfigure}
     
     \vspace{5mm}
     \centering
     \begin{subfigure}[b]{0.29\textwidth}
        \centering
        \includegraphics[width=\textwidth]{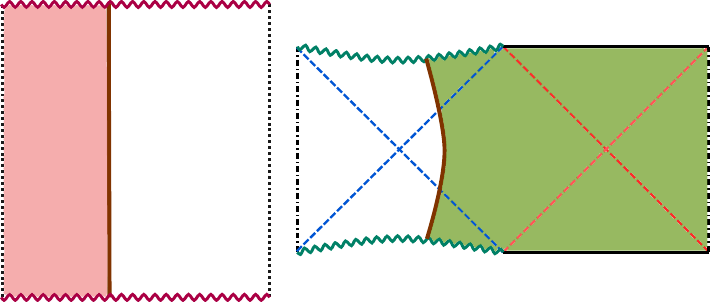}
        \caption*{(D1a)}
        \label{fig:D1a}
     \end{subfigure}
     \hfill
     \centering
     \begin{subfigure}[b]{0.29\textwidth}
        \centering
        \includegraphics[width=\textwidth]{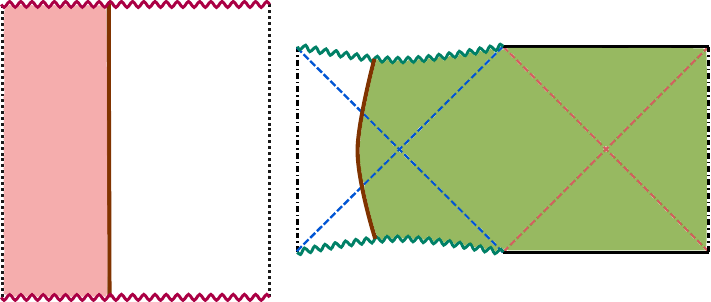}
        \caption*{(D1b)}
        \label{fig:D1b}
     \end{subfigure}
     \hfill
     \centering
     \begin{subfigure}[b]{0.29\textwidth}
        \centering
        \includegraphics[width=\textwidth]{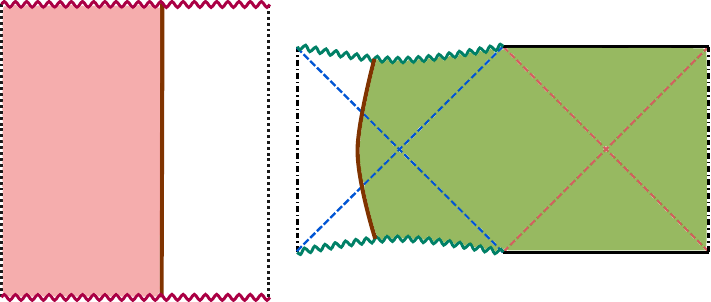}
        \caption*{(D2)}
        \label{fig:D2}
     \end{subfigure}
     
     \vspace{5mm}
     \centering
     \begin{subfigure}[b]{0.29\textwidth}
        \centering
        \includegraphics[width=\textwidth]{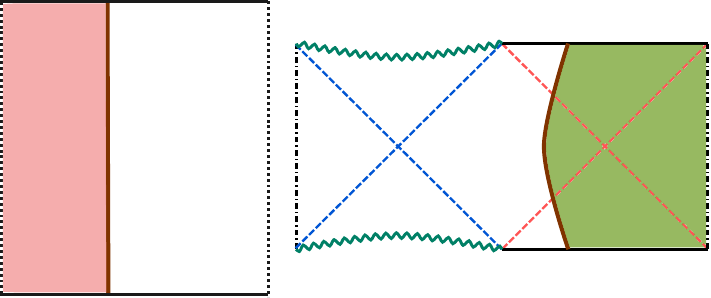}
        \caption*{(D3a)}
        \label{fig:D3a}
     \end{subfigure}
     \hfill
     \centering
     \begin{subfigure}[b]{0.29\textwidth}
        \centering
        \includegraphics[width=\textwidth]{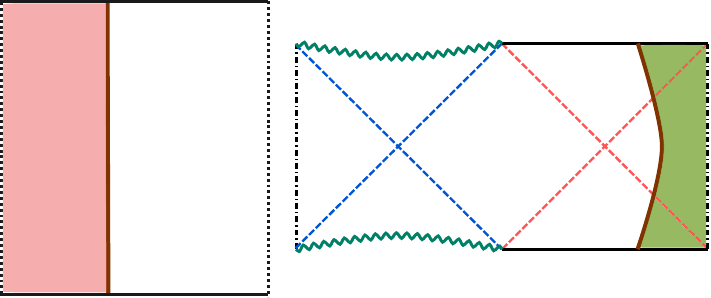}
        \caption*{(D3b)}
        \label{fig:D3b}
     \end{subfigure}
     \hfill
     \centering
     \begin{subfigure}[b]{0.29\textwidth}
        \centering
        \includegraphics[width=\textwidth]{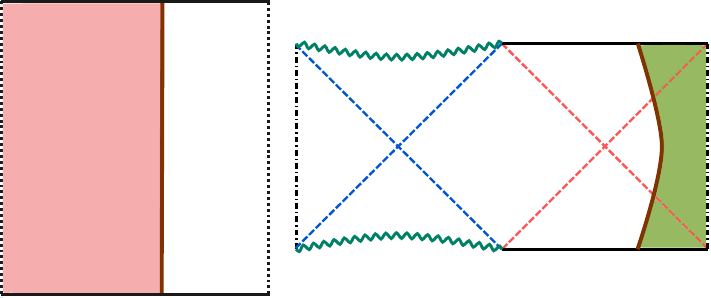}
        \caption*{(D4)}
        \label{fig:D4}
     \end{subfigure}
     \hfill

    \caption{Possible time reversal symmetric thin shell spacetimes}
    \label{fig:TimeSym}
\end{figure}

\begin{figure}
     \centering
     \hspace{0.1\textwidth}
     \begin{subfigure}[b]{0.3\textwidth}
        \centering
        \includegraphics[width=\textwidth]{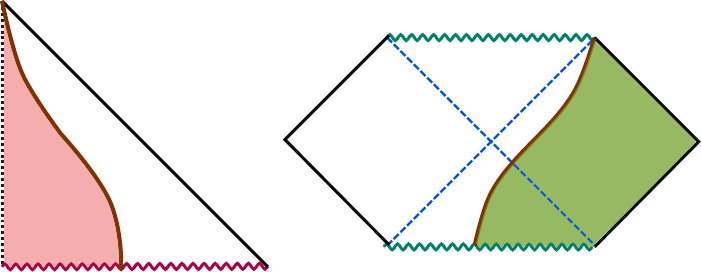}
        \caption*{(M3)}
        \label{fig:M3}
     \end{subfigure}
     \hspace{0.1\textwidth}
     \centering
    \begin{subfigure}[b]{0.3\textwidth}
         \centering
         \includegraphics[width=\textwidth]{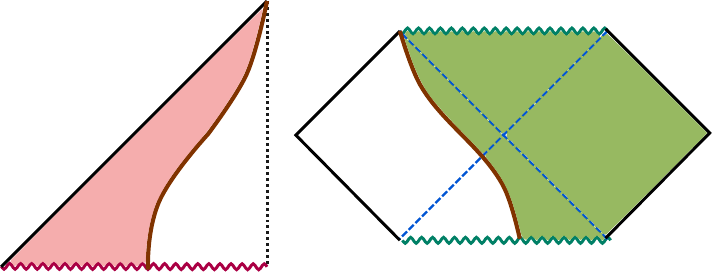}
         \caption*{(M4)}
         \label{fig:M4}
     \end{subfigure}
    \hspace{0.1\textwidth}
          
      \vspace{5mm}
      \centering
       \hspace{0.1\textwidth}
      \begin{subfigure}[b]{0.3\textwidth}
         \centering
         \includegraphics[width=\textwidth]{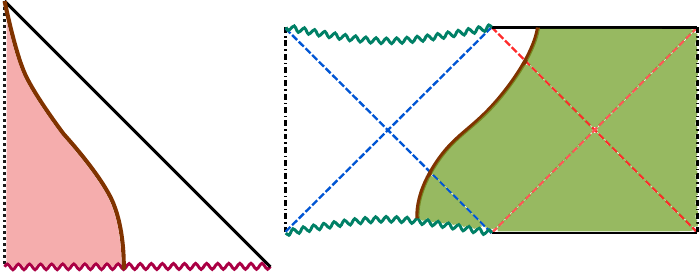}
          \caption*{(D5)}
         \label{fig:D5}
     \end{subfigure}
      \hspace{0.1\textwidth}
     \centering
     \begin{subfigure}[b]{0.3\textwidth}
         \centering
        \includegraphics[width=\textwidth]{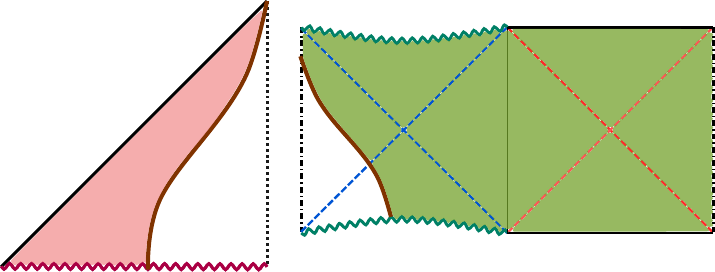}
         \caption*{(D6)}
         \label{fig:D6}
     \end{subfigure}
       \hspace{0.1\textwidth}
     
     \vspace{5mm}
     \centering
     \begin{subfigure}[b]{0.3\textwidth}
        \centering
        \includegraphics[width=\textwidth]{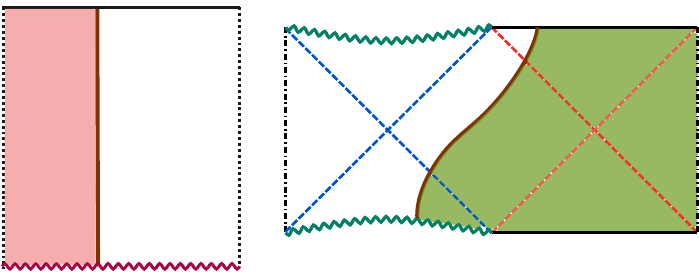}
        \caption*{(D7a)}
        \label{fig:D7a}
     \end{subfigure}
     \hfill
     \centering
     \begin{subfigure}[b]{0.3\textwidth}
        \centering
        \includegraphics[width=\textwidth]{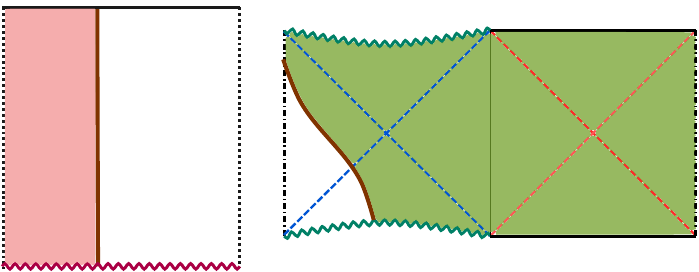}
        \caption*{(D7b)}
        \label{fig:D7b}
     \end{subfigure}
     \hfill
     \centering
     \begin{subfigure}[b]{0.3\textwidth}
        \centering
        \includegraphics[width=\textwidth]{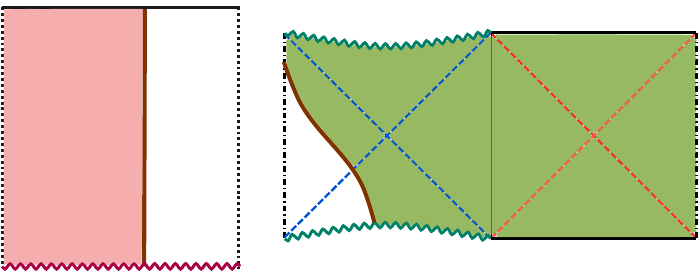}
        \caption*{(D8)}
        \label{fig:D8}
     \end{subfigure}

 \caption{Possible solutions with a monotonically expanding shell}
    \label{fig:Mono}
\end{figure}

\begin{table}[h]
\centering
\renewcommand{\arraystretch}{1.1}
\begin{tabular}{c c c c c c}
\hline
Label & $\Lambda$ & FLRW type & Geometry & $\beta$ sign & Time symmetry \\
\hline
A1a & $<0$ & Open (BB--BC) & Bubble & $+$ & Yes \\
A1b & $<0$ & Open (BB--BC) & Bubble & $-$ & Yes \\
A2  & $<0$ & Open (BB--BC) & Swiss cheese & $+$ & Yes \\
A3a & $<0$ & Closed ($<\tfrac12$ sphere, BB--BC) & Bubble & $+$ & Yes \\
A3b & $<0$ & Closed ($<\tfrac12$ sphere, BB--BC) & Bubble & $-$ & Yes \\
A4  & $<0$ & Closed ($>\tfrac12$ sphere, BB--BC) & Bubble & $-$ & Yes \\
\hline
M1a & $=0$ & Closed ($<\tfrac12$ sphere, BB--BC) & Bubble & $+$ & Yes \\
M1b & $=0$ & Closed ($<\tfrac12$ sphere, BB--BC) & Bubble & $-$ & Yes \\
M2  & $=0$ & Closed ($>\tfrac12$ sphere, BB--BC) & Bubble & $-$ & Yes \\
\hline
D1a & $>0$ & Closed ($<\tfrac12$ sphere, BB--BC) & Bubble & $+$ & Yes \\
D1b & $>0$ & Closed ($<\tfrac12$ sphere, BB--BC) & Bubble & $-$ & Yes \\
D2  & $>0$ & Closed ($>\tfrac12$ sphere, BB--BC) & Bubble & $-$ & Yes \\
D3a & $>0$ & Closed (bounce) & Bubble & $+$ & Yes \\
D3b & $>0$ & Closed (bounce) & Bubble & $-$ & Yes \\
D4  & $>0$ & Closed (bounce, $>\tfrac12$ sphere) & Bubble & $-$ & Yes \\
\hline
M3  & $=0$ & Open (BB) & Bubble & $+$ & No \\
M4  & $=0$ & Open (BB) & Swiss cheese & $-$ & No \\
\hline
D5  & $>0$ & Open (BB) & Bubble & $+$ & No \\
D6  & $>0$ & Open (BB) & Swiss cheese & $-$ & No \\
D7a & $>0$ & Closed ($<\tfrac12$ sphere, BB) & Bubble & $+$ & No \\
D7b & $>0$ & Closed ($<\tfrac12$ sphere, BB) & Bubble & $-$ & No \\
D8  & $>0$ & Closed ($>\tfrac12$ sphere, BB) & Bubble & $-$ & No \\
\hline
\end{tabular}
\caption{Classification of thin-shell spacetimes shown in Figs.~5 and~6. 
Here `BB--BC' denotes a big-bang/big-crunch cosmology, while `BB' denotes a FLRW cosmology that begins at a big-bang singularity. 
`Bounce' denotes a time-reversal symmetric FLRW cosmology with a non-zero minimum scale factor. 
`Bubble' denotes a finite FLRW region surrounded by a Schwarzschild exterior, whereas `Swiss cheese' denotes a Schwarzschild vacuole embedded in an infinite FLRW spacetime. 
The sign of $\beta$ determines whether the shell lies outside ($\beta>0$) or entirely behind ($\beta<0$) the black-hole horizon at the reference time slice.}
\label{table:Catalog}
\end{table}

In this section, we describe and categorise the causal structures of all the possible types of thin shell spacetimes with a shell of dust outlined in \cref{subsec:SimpleState}.
The causal structures of the empty shell solutions are identical and can be represented by the same Penrose diagrams.

\subsubsection*{Classification of solutions}
Each panel in \cref{fig:TimeSym} and \cref{fig:Mono} shows the Penrose diagram of a cosmological spacetime and a black hole spacetime, marked with the trajectory of the thin shell. 
The shaded patches in the FLRW spacetime and the black hole spacetime are glued along the shell to produce a unique thin shell geometry. 

The solutions are categorised according to the cosmological constant of the spacetime (represented by capital letters A for AdS, D for dS, and M for Minkowski), and the type of the cosmological patch (labelled by numbers $1,2,\dots$ ). 
Some of these solutions (for example A1a and A1b) have the same type of cosmological patch, but differ by the sign of $\beta$ (defined in Eq. \eqref{eq:betaFormula}). 
Solutions of type `a' have $\beta>0$ and the shell is outside the black hole horizon for some time the perspective of a Schwarzschild observer. Solutions of type `b' have $\beta<0$ and the shell is completely behind the horizon.
The full classification of these solutions is tabulated in \cref{table:Catalog}.
{
The different solutions catalogued here have different causal structures as is clear from \cref{fig:Mono} and \cref{fig:TimeSym}, and hence are distinct solutions unrelated by coordinate transformations.}

In \cref{sec:Cosmoshell} we showed that the proper size of the shell is proportional to the scale factor.
Therefore, the various types of thin shell spacetimes are related to different classes of scale factors; time symmetric scale factors give rise to time-symmetric thin shell spacetimes and monotonic scale factors give rise to monotonic geometries.

\subsubsection*{Time-reversal symmetric solutions}
In \cref{fig:TimeSym} we list all the time-reversal symmetric thin shell geometries.
In spacetimes A1-A4, M1, M2, D1, and  D2 we have a FLRW with big bang/big crunch singularities. This implies that the shell starts and ends with zero size at the big bang and big crunch singularity respectively, completing the journey in finite proper time. 
In the Schwarzschild geometry, the bubble surface also starts at the past singularity and ends at the future singularity of the black hole, following a timelike trajectory. 
Thus in the composite spacetime, the black hole singularities and the FLRW singularities join up.

We can also have solutions involving big bounce cosmologies, as shown in cases D3 and D4.
In this scenario, the shell has a minimum size at the time-symmetric slice and it grows out to infinite proper size in the past and the future.
On the black hole side, the bubble surface starts at the past infinity and ends at the future infinity following a timelike trajectory, after crossing the cosmological horizon twice. 

\subsubsection*{Solutions with monotonic scale factors}
In \cref{fig:Mono} we show possible scenarios involving big bang cosmologies.
Here, the shell starts with zero size and continues to expand monotonically.
Note that these constructions are possible only in backgrounds with zero $\lambda$ (as shown in cases M3, M4) or positive $\lambda$ (cases D5-D8).
From the perspective of the black hole spacetime, the shell starts at the past singularity and emerges out of the event horizon, before meeting the future asymptotic infinity. 
In this case also the singularities and infinities of both the spacetimes are identified. 

\subsubsection*{Bubbles of cosmology and Swiss cheese spacetimes}
In the case of open FLRWs, the cosmological patch of the thin shell geometries can be of two types. 
In spacetimes A1, M3, and D5 we have a bubble of FLRW spacetime that covers a finite region $r<r_0$ in terms of the comoving radius.  
Alternatively, we may have instead a bubble of black hole spacetime within a portion of the FLRW spacetime that extends over the region $r>r_0$, as seen in cases A2, M4, and D6.

In the latter case, the FLRW region remains homogeneous away from the black hole bubble; an observer here is oblivious to the presence of a black hole bubble. Thus, we can introduce several such bubbles, with possibly different sizes and holding black holes of different masses, as long as they do not overlap. 
At any time slice of the full spacetime resembles a slice of `Swiss cheese' (\cref{fig:SwissSlice}), with homogeneous FLRW `cheese' filling up the region between pockets of black holes.
The bubble of cosmology spacetime, however, can only hold a single bubble, as introducing any more bubbles will break the spherical symmetry of the Schwarzschild spacetime. 

In thin shell geometries involving a closed FLRW, the cosmological patch is a ball shaped subregion of a sphere. In cases A3, M1, D1, D3, and D7 the patch covers less than half of the full sphere and in cases A4, M2, D2, D4, and D8 the patch covers more than half of the full sphere.

\subsection{Parameter regimes}
\label{sec:parameter}
In this section, we study the parameter regimes for the validity of the solutions listed in the previous section. 
For concreteness, we will discuss only the case of dust and radiation filled FLRWs in 3+1 dimensions. 

\subsubsection*{Time-reversal symmetric solutions}
Let us consider a case with time-reversal symmetric scale factors.
We can always shift the origin of the time coordinate such that 
$t=0, T(0)=0$ correspond to the time-reversal symmetric slice with $H(0)=0$.
We can also set $a(0)=1$ without loss of generality since the Hubble's constant does not change under scaling of $a(t)$.
The spatial curvature of the FLRW is fixed as
\begin{equation}
    \K=\rho_M + \rho_R + \lambda,
\end{equation}
using the Friedmann equation.
The sign of $\Ddot{a}(0)$ determines whether the scale factor grows away from the time symmetric slice. 
We have a big bounce cosmology if and only if $\ddot{a}(0) > 0$ which is equivalent to the condition 
\begin{equation}
    \rho_M + 2\rho_R < 2\lambda.
\end{equation}
In all other cases, we have a big bang/big crunch cosmology. 

\paragraph{Bubbles of cosmology}
A class of allowed geometries are those 
with a finite bubble of FLRW ($r<r_0$) enclosed within a Schwarzschild exterior. 
To construct these bubbles of cosmology spacetimes we choose $q=1$.
Then Eq. \eqref{eq:muDustAndRad} gives us the mass of the black hole
\begin{equation}
\label{eq:BHMassBubble}
   \mu = \rho_M R_0^3 \pm 2\sqrt{\rho_R}R_0^2 \sqrt{1 - R_0^2(\rho_M + \rho_R + \lambda)},
\end{equation}
where the minus sign is to be taken only while considering closed FLRWs with more than half of the full spatial slice included. 
Notice that in this case we also have a lower bound
 \[R_0 > 2 \sqrt{\rho_R}/\sqrt{(\rho_M + 2 \rho_R)^2 + 4 \lambda \rho_R}\] for
the black hole mass to be positive\footnote{At this critical value of $R_0$ the solution degenerates into two disconnected parts, the exterior spacetime becoming a pure AdS, Minkowski or dS spacetime,  and the interior spacetime becoming a closed universe with a bubble of FLRW spacetime glued together with a bubble of maximally symmetric spacetime}.

We can also check that at the time-reversal symmetric slice the bubble is in the exterior of the black hole (or between the cosmological and black hole horizon in the case of the Schwarzschild-de Sitter black hole). 
To see this it is sufficient to show that $F(R_0)$ is positive. 
Using the black hole mass from Eq. \eqref{eq:BHMassBubble}, we see $F(R_0) = (\sqrt{\rho_R} R_0 +R'_\K(r_0))^2$, which is clearly positive. 

Since the surface of the bubble is in the black hole exterior, we need to determine whether the bubble is completely behind the horizon for a Schwarzschild observer, i.e. if the retained black hole spacetime contains the horizon at the time-reversal symmetric slice. 
The bubble will be behind the horizon if and only if $\beta(t=0) < 0$, or equivalently $Q=-1$.
Considering Equation \eqref{eq:JC2applied} at $t=0$, we see that this is equivalent to
\begin{equation}
\label{eq:bubBehindHorizon}
{R_{\K}'(r_0) } < \sqrt{\rho_R} R_{\K}(r_0) \; .
\end{equation}
This will be true if and only if
\begin{equation*}
   R'_\K(r_0) < 0 \qquad \mathrm{or}, \qquad 
R_0 > 1/\sqrt{2 \rho_R + \rho_M + \lambda} \; .
\end{equation*}
We have $Q=1$ and the bubble is outside the horizon at the time-symmetric slice whenever the above condition fails, i.e. if
\begin{equation*}
   R'_\K(r_0) > 0 \qquad \mathrm{and}, \qquad 
R_0 < 1/\sqrt{2 \rho_R + \rho_M + \lambda} \; .
\end{equation*}

\paragraph{Swiss cheese}
To construct Swiss cheese geometries we start by choosing $q=-1$ which gives 
gives us black hole mass,
\begin{equation}
\label{eq:BHMassBubble_swiss}
   \mu = \rho_M R_0^3 \mp 2\sqrt{\rho_R}R_0^2 \sqrt{1 - R_0^2(\rho_M + \rho_R + \lambda)}.
\end{equation}
This time, the plus sign corresponds to the choice $R'_K(r_0) < 0$.
Thus while considering open FLRWs or closed FLRWs with less than half of the full space included, we have a minimum vacuole size of 
\begin{equation*}
    R_0 \geq \frac{2\sqrt{\rho_R}}{\sqrt{(\rho_M+ 2\rho_R)^2 + 4\rho_R\lambda}}.
\end{equation*}
Similar to the previous case, the shell is in the exterior of the black hole at $t=0$.
Then on this time slice either the event horizon is included along ($\beta < 0$) or the horizon is excised and the black hole spacetime extends over infinite proper size ($\beta > 0$).
Applying the second junction condition Eq. \eqref{eq:JC2applied} at $t=0$, we see $\beta = -R'_K(r_0) - \sqrt{\rho_R}R_0$.
$\beta$ is positive if and only if 
\begin{equation*}
   R'_\K(r_0) < 0 \qquad \mathrm{and,} \qquad R_0 < 1/\sqrt{2\rho_R + \rho_M + \lambda}.
\end{equation*}
In all other cases, $\beta$ is negative and hence the black hole spacetime includes the event horizon.

\subsubsection*{Monotonically expanding shells}
In such a case it is most natural to fix $a=1$ at the moment the shell just emerges out of the event horizon and specify any initial conditions here. 
The proper size of the shell $R_0$ at this point is by definition equal to the horizon radius and hence the black hole mass can be found by equating $F(R_0)=0$, giving us $\mu = R_0 - \lambda R_0^{3}$.
Since we want $R_0$ to correspond to the black hole horizon we have the condition 
\begin{equation}
    R_0 < \sqrt{\frac{1}{3\lambda}}.
\end{equation}
However, since the black hole mass is also given by Eq. \eqref{eq:muDustAndRad}, 
equating the expression for $\mu$ obtained in these two ways we obtain
\begin{align}
\label{eq:RKprime_Bang}
R'_\K(r_0) &= \frac{1-(\lambda+\rho_M)R_0^2}{2q\sqrt{\rho_R}R_0 }.
\end{align}
Recalling that $R'_\K(r_0)^2 = 1- \K R_0^2$, we obtain the curvature of the FLRW 
\begin{equation}
    \K = \frac{1}{R_0^2} - \frac{1}{4\rho_R}\left(\frac{1}{R_0^2} - \rho_M - \lambda\right)^2.
\end{equation}
Since $\lambda \geq 0$ and $\K<\K_c$ for big bang cosmologies the allowed values of $R_0$ are given by
\begin{align}
\label{eq:R0allowed1}
    R_0^2 &< \frac{1}{{\rho_M+2\rho_R+\lambda + \sqrt{(\rho_M+2\rho_R+\lambda)^2-(\rho_M+\lambda)^2 - 4\rho_R\K_c}}} \quad \mathrm{or,} \\
\label{eq:R0allowed2}
    R_0^2 &> \frac{1}{{\rho_M+2\rho_R+\lambda -\sqrt{(\rho_M+2\rho_R+\lambda)^2-(\rho_M+\lambda)^2 - 4\rho_R\K_c}}} .
\end{align}
Similar to previous cases, whether the black hole horizon is included in the full spacetime is determined by the sign of $\beta = qR'_K(r_0)-\sqrt{\rho_R}R_0$. It is straightforward to check that $\beta > 0$ if and only if $R_0^2 < 1/(\rho_M+2\rho_R+\lambda )$. 
Thus full spacetime satisfying Eq. \eqref{eq:R0allowed1} have $\beta>0$ and its black hole side contains the region outside the event horizon. 
Conversely, the black hole part of the spacetimes satisfying Eq. \eqref{eq:R0allowed2} contains the region inside the horizon. 
Note that our analysis did not depend on the sign of $q$. Thus the same conclusions are applicable for bubble of cosmology spacetimes with $q=1$ and Swiss cheese spacetimes with $q=-1$.

\section{Applications and generalisations}

\paragraph{Applications}
Owing to the large parameter space associated with dust- and/or radiation-filled cosmologies, our construction gives rise to a wide variety of distinct spacetime geometries, as illustrated in \cref{fig:TimeSym,fig:Mono}.
Each solution corresponds to a particular choice of parameters and global structure.
Some of these spacetimes describe a finite bubble of cosmology embedded within a black hole, while others resemble Swiss-cheese geometries, consisting of one or more vacuum regions interspersed within a matter-filled cosmological background.
The bubble solutions provide idealised examples of black-hole formation through gravitational collapse and have also been interpreted as semiclassical black-hole microstates in related contexts \cite{Balasubramanian_2022_Microscopic,Balasubramanian_2022_microscopicGR,Emparan_2024_Microstates}.
Swiss-cheese–type solutions, on the other hand, furnish analytic examples of inhomogeneous spacetimes that nevertheless evolve according to the FLRW equations on large scales.
Such geometries have long been used to model structure formation in an expanding universe \cite{Einstein_1945_influence} and to explore alternative explanations of cosmological acceleration driven by inhomogeneities \cite{Marra_2007_SwissCheese,Vanderveld_2008_LuminosityI,Vanderveld_2008_LuminosityII}.

{
A significant subset of the solutions constructed here is time-reversal symmetric, admitting a well-defined Euclidean continuation about the time-symmetric slice.
In separate work by the author and collaborators \cite{Sahu2023Bubbles}, this Euclidean continuation was used to argue for the existence of a holographic dual description for cosmological bubbles with $\lambda<0$ embedded in an AdS--Schwarzschild background.
In that analysis, suitable Euclidean boundary conditions allow the preparation of a dual quantum state within the AdS/CFT correspondence.
The present paper does not develop this holographic interpretation further, but instead provides the Lorentzian thin-shell geometries that underlie such constructions.
Related holographic approaches to big-bang/big-crunch cosmologies with negative cosmological constant include models based on thin shells or end-of-the-world branes as low-energy effective descriptions \cite{cooper_black_2018,Antonini2019,Waddell_2022_Bottom-up,Fallows2022Constraints,Ross2022Cosmologies,Antonini_2023_cosmology_random_entanglement}.
It would be interesting to explore whether the $\lambda \geq 0$ spacetimes discussed here admit analogous interpretations in the context of asymptotically flat \cite{Pasterski_2016_FlatSpace} or de Sitter holography \cite{Strominger:2001pn}.

It is worth noting that related Swiss-cheese–type cosmologies arise in lower-dimensional settings as well.
In particular, in \cite{SahuBHcosmo} the author studied cosmological spacetimes in $2+1$ dimensions in which the matter content consists of a uniform lattice of black holes.
These geometries admit a dual state prepared via a Euclidean path integral over a hyperbolic handlebody, providing another example in which inhomogeneous cosmologies admit a controlled semiclassical and holographic description.

Thin-shell techniques are also widely employed in stellar modelling.
The classic example is the Oppenheimer--Snyder solution, in which a homogeneous dust interior is matched to an exterior Schwarzschild spacetime.
More general stellar models involve matching regions with different equations of state across timelike hypersurfaces, with the junction conditions determining the motion and stress-energy of stellar surfaces or internal interfaces.
From this perspective, the cosmological thin shells constructed here may be viewed as a complementary family of exact solutions in which the interior region is an FLRW cosmology and the shell plays the role of an idealised surface separating matter from vacuum.}

It is unlikely that the dust-shell solutions presented here are realised in realistic astrophysical settings, and we are not aware of any physical mechanism that would produce stars surrounded by pressureless dust layers.
Moreover, such configurations may be unstable under non-spherical perturbations, for instance through Rayleigh--Taylor instabilities that cause the dust to clump and penetrate the cosmological region.\footnote{The author thanks William Unruh for pointing this out.}
Nevertheless, as exact solutions of Einstein’s equations, these spacetimes provide a useful theoretical laboratory for studying cosmological matching problems and for exploring conceptual aspects of quantum cosmology, particularly in connection with holography.

\paragraph{Possible generalisations}
Our choice to consider a shell whose stress-energy tensor is conserved on the
worldvolume,
\begin{equation}
S^{a}{}_{b|a} = 0,
\end{equation}
was made primarily for simplicity.
As shown in section~3, this assumption implies that the shell moves along the
streamlines of the cosmological fluid, i.e.\ it is comoving with the FLRW matter.
Equivalently, the shell neither exchanges energy nor momentum with the cosmological region.
From the point of view of the cosmology, the shell acts as a reflecting boundary in the sense that the matter distribution continues to evolve homogeneously and isotropically.

{
More general configurations with $S^{a}{}_{b|a} \neq 0$ are certainly possible in principle.
In such cases the thin shell would exchange energy and momentum with the cosmological region, and the stress--energy flux across the hypersurface would be non-zero.
One way to model this situation would be to include interaction terms in the matter sector of the gravitational effective action \cref{eq:gravAction} that couple the cosmological matter to degrees of freedom living on the shell, while still requiring the exterior spacetime to be vacuum or black hole–like.
The junction conditions would then need to be supplemented by additional dynamical equations governing the exchange between the shell and the cosmological region, subject to the requirement that no energy dissipates into the external vacuum.
A detailed analysis of such non-conserved shells would require a more microscopic model of the physics at the interface and lies beyond the scope of the present work.

Another interesting generalisation would be to allow the cosmological matter to dissipate through the shell into the exterior region.
For instance, if the cosmological region contains radiation that leaks outward, the exterior geometry would no longer be vacuum but could instead be described by a Vaidya spacetime with outgoing null radiation.
In this case one may consider a gluing problem that joins an FLRW cosmological bubble to an external Vaidya region.
Related thin-shell constructions have been used in the context of radiating stellar collapse \cite{Santos:1985inc,Santos2}, and it would be interesting to adapt similar methods to the cosmological setting.

While in this paper we restricted attention to cosmological spacetimes, the framework developed here applies more generally to any matter-filled spacetime separated from a vacuum region by a hypersurface.
In such cases the condition $T_{\alpha\beta} e^\alpha_a n^\beta = 0$ ensures that there is no net flux of stress-energy across the surface, and the shell stress-energy can again be determined from the junction conditions.
A relation analogous to \cref{eq:shellfromBulk} may then be derived for more general bulk matter configurations.

Finally, in our analysis the cosmological region was taken to be a spherically symmetric portion of an FLRW universe, which is both homogeneous and isotropic.
One may instead consider a spherically symmetric bubble of Lemaître--Tolman--Bondi spacetime, which preserves isotropy about the centre but allows for radial inhomogeneity in the matter distribution.
It would be interesting to repeat the analysis of \cref{sec:Cosmoshell} in this more general setting, and to understand how the Gauss--Codazzi constraints and junction conditions are modified in the presence of inhomogeneous cosmological matter.
We leave this investigation for future work.

All results in this work are derived within Einstein gravity.
In modified theories such as $f(R)$ gravity the junction conditions are altered by additional continuity requirements and surface contributions, which we do not consider here.
For discussions of thin-shell junction conditions in $f(R)$ gravity we refer the reader to \cite{Senovilla:2013vra,KUMAR2024116690}.
}

\section{Discussion}

\paragraph{Summary and salient features}
We have discussed the formulation of junction conditions at a surface separating a vacuum spacetime
from a general homogeneous and isotropic cosmological spacetime. 
The separating surface may not be empty; instead, it may be a physical shell with a stress-energy tensor related to the discontinuity of the extrinsic curvature of the surface.
{
A key lesson of this analysis is that, once matter is present, Israel’s junction conditions must be supplemented by the constraints imposed by the Gauss–Codazzi equations.
When both sides of the hypersurface are vacuum these constraints are automatically satisfied, and the matching reduces to solving the junction conditions alone.
In contrast, for a cosmological region with nonzero matter pressure, the constraints are nontrivial because generic hypersurfaces carry stress-energy flux and disturb homogeneity.
Our approach treats the Gauss–Codazzi constraints as part of the matching problem.
In particular, imposing conservation of the shell stress-energy forces the shell to be comoving with the FLRW fluid, ensuring vanishing flux across the hypersurface and preserving the homogeneous and isotropic evolution of the cosmology.
}

As a concrete example, we considered cosmologies filled with uniform pressureless dust and/or radiation and found two classes of solutions in which the shell matter takes a simple and controlled form, namely ideal fluids confined to the shell.
One of these solutions, valid in any $D+1$ dimensional spacetime with $D \geq 3$, reproduces the well-known Oppenheimer–Snyder construction, in which the cosmological region is pressureless and the separating hypersurface carries no stress-energy.
The second class of solutions is special to $3+1$ dimensional spacetime and involves a dust- and radiation-filled cosmology matched to vacuum by a thin shell of pressureless matter whose surface density is proportional to the square root of the radiation density.

{
As explained in \cref{subsec:SimpleState}, the special role of $3+1$ dimensions arises from the combinatorics of matching various scale-factor dependent terms appearing in the junction conditions, while requiring a simple mixture of ideal fluids as the shell matter content.
Only in four spacetime dimensions can the contributions sourced by bulk radiation be balanced by a pressureless shell while maintaining energy conservation on the shell.
In higher dimensions the same matching problem can be solved, but generically requires shell matter with a nontrivial equation of state.
More generally, once the bulk FLRW matter content is specified, the required shell stress-energy can be determined directly from the junction conditions.
In particular, \cref{eq:shellfromBulk} provides a closed-form expression for the surface energy density needed to sustain an FLRW bubble with arbitrary homogeneous and isotropic bulk matter.
The dust- and radiation-filled cosmology in $3+1$ dimensions discussed above is a distinguished special case of this general framework, in which the required shell matter simplifies to pressureless dust.
}

These constructions exist for a variety of FLRW cosmologies with different spatial topologies (open or closed),
distinct causal structures (time-reversal symmetric big-bang/big-crunch or bouncing cosmologies, as well as monotonically expanding cosmologies with an initial singularity),
and different signs of the cosmological constant.
Altogether, this leads to a family of twenty-two distinct thin-shell spacetimes.
Their classification, including the sign of the cosmological constant, the nature of the FLRW region, and whether the geometry represents a bubble of cosmology or a Swiss-cheese configuration, is summarised in \cref{table:Catalog}.


\paragraph{Relation with earlier models}
{
These results also clarify the relation to earlier collapse and thin-shell constructions.
The Oppenheimer--Snyder model occupies a special corner of the present framework in which the interior cosmological matter has vanishing pressure.
In this regime, the Gauss--Codazzi constraints impose no additional restrictions beyond Israel’s junction conditions, and matching a cosmological interior to a vacuum exterior is straightforward.
Within our parameter space, the Oppenheimer--Snyder solution is recovered smoothly as the limit of the dust--radiation family in which the radiation density is taken to zero.

Once bulk pressure is present, however, the situation changes qualitatively.
Generic hypersurfaces in a cosmological spacetime then carry stress--energy flux and disturb the homogeneous and isotropic evolution, obstructing a naive matching to vacuum.
The four-dimensional dust-shell solution identified here demonstrates that this obstruction can nevertheless be resolved in a controlled manner.
For a comoving shell obeying surface stress--energy conservation, the junction conditions force the shell to acquire pressureless dust whose energy density is uniquely fixed by the radiation content of the cosmology, rather than introduced as an independent assumption.

Within the class of spherically symmetric thin-shell spacetimes with an FLRW interior and a vacuum exterior, this provides an explicit example in which a cosmological region with non-zero fluid pressure can be consistently matched to vacuum without inducing shell pressure or energy exchange.
}

\section*{Acknowledgements}
The author would like to thank Mark Van Raamsdonk and Petar Simidzija for early collaboration and comments on the draft, and William G. Unruh for useful discussions. 
The author is supported by a Four-Year Doctoral Fellowship from the University of British Columbia.

\bibliographystyle{jhep}
\bibliography{references}

\end{document}